\documentclass[12pt,preprint]{aastex}

\newcommand{\etal}{et~al.}
\newcommand{\fig}{Fig.~}
\newcommand{\mic}{\,$\mu$m}
\newcommand{\pc}{\,pc}
\newcommand{\av}{A_V}
\newcommand{\ak}{A_K}

\newcommand{\plam}{p_\lambda}
\newcommand{\lmax}{\lambda_{\rm max}}
\newcommand{\pmax}{p_{\rm max}}
\newcommand{\taulam}{\tau_\lambda}
\newcommand{\tauk}{\tau_K}
\newcommand{\ratlam}{p_\lambda/\tau_\lambda}
\newcommand{\ratk}{p_K/\tau_K}

\newcommand{\evk}{E_{V-K}}
\newcommand{\ejk}{E_{J-K}}
\newcommand{\ejh}{E_{J-H}}
\newcommand{\ehk}{E_{H-K}}
\newcommand{\water}{H$_2$O}
\newcommand{\oph}{$\rho$~Oph}

\slugcomment{Version date: \today}

\shorttitle{CO$_2$ paper}
\shortauthors{Whittet et al.}

\begin{document}

\title{THE EFFICIENCY OF GRAIN ALIGNMENT IN DENSE INTERSTELLAR CLOUDS: 
		A REASSESSMENT OF CONSTRAINTS FROM NEAR INFRARED POLARIZATION\\~}

\author{D.\ C.\ B. Whittet$^{1}$, J.\ H. Hough$^{2}$, A. Lazarian$^{3}$ \& Thiem Hoang$^{3}$\\~}

\altaffiltext{1}{Department of Physics, Applied Physics \& Astronomy,
	Rensselaer Polytechnic Institute, 110 Eighth Street, Troy, NY 12180. Email: whittd@rpi.edu.}

\altaffiltext{2}{Centre for Astrophysics Research, University of Hertfordshire,
	Hatfield, AL10 9AB, U.K.}

\altaffiltext{3}{Department of Astronomy, University of Wisconsin,
	Madison, WI 53706.}

\begin{abstract}
We present the results of a detailed study of interstellar polarization efficiency (as measured by the ratio $\ratlam$) toward molecular clouds, with the aim of discriminating between grain alignment mechanisms in dense regions of the interstellar medium. The data set includes both continuum measurements in the $K$ (2.2\mic) passband and values based on ice and silicate spectral features. Background field stars are used to probe polarization efficiency in quiescent regions of dark clouds, yielding a dependence on visual extinction well-represented by a power law ($\ratlam \propto [\av]^{-0.52}$), in agreement with previous work. No significant change in this behavior is observed in the transition region between the diffuse outer layers and dense inner regions of clouds, where icy mantles are formed, and we conclude that mantle formation has little or no effect on the efficiency of grain alignment. The field-star data are used as a template for comparison with results for embedded young stellar objects (YSOs). The latter generally exhibit greater polarization efficiency compared with field stars at comparable extinctions, some displaying enhancements in $\ratlam$ by factors of up to $\sim 6$ with respect to the power-law fit. Of the proposed alignment mechanisms, that based on radiative torques appears best able to explain the data. The attenuated external radiation field appears adequate to account for the observed polarization in quiescent regions for extinctions up to $\av\sim 10$~mag. Radiation from the embedded stars themselves may enhance alignment in the lines of sight to YSOs. Enhancements in $\ratlam$ observed in the ice features toward several YSOs are of greatest significance, as they demonstrate efficient alignment in cold molecular clouds associated with star formation. We advocate spectropolarimetry of ice features in background stars as a reliable means of mapping internal magnetic fields within molecular clouds. To date, such observations exist for just one line of sight.

\end{abstract}

\keywords{polarization --- dust, extinction --- ISM: magnetic fields --- infrared: ISM}

\section{Introduction}
The detailed physics of grain alignment is yet to be fully understood despite much activity and considerable progress in the years since the pioneering work of Davis \& Greenstein (1951; DG) (see reviews by Lazarian 2003, 2007). Developing a comprehensive, realistic model for grain aligment is an important astrophysical goal for several reasons. Magnetic fields play vital roles in interstellar processes and star formation, and the polarization that results from grain alignment provides the means to study magnetic fields if the alignment process is well understood. Polarization is also a valuable tool for studying the dust grains themselves, placing constraints on their magnetic properties, structure, shape and size distribution. It is generally assumed that the dominant process responsible for alignment involves interaction between the spin of the particles and the ambient magnetic field, as originally proposed by DG: a spinning grain tends to become orientated with its longest axis perpendicular to the angular momentum vector, and paramagnetic relaxation imposes alignment of the angular momentum with respect to the magnetic field vector, ${\bf B}$. For polarization produced by dichroic extinction, the mean direction of the electric vector in the transmitted beam is thus parallel to ${\bf B}$, and so, in principle, the observed polarization traces the magnetic field in the plane of the sky. In practise, the observed polarization depends on the mean field direction, weighted according to the distribution of dust along the line of sight and the efficiency with which it is aligned. 

The original DG formulation assumed the grains to be spinning thermally and to have paramagnetic properties (see Roberge \& Lazarian 1999 for a modern account). This mechanism was qualitatively successful, explaining the pattern of interstellar polarization observed in the Milky Way, but it failed quantitatively to predict significant alignment under realistic physical conditions. Subsequent investigations considered enhancements to the DG mechanism that might arise if the grains are endowed with superparamagnetic properties or if they spin suprathermally. The Purcell (1979) mechanism attributes suprathermal spin primarily to H$_2$ formation at active sites on the grain surfaces. More recent work has emphasized radiative torques (RT; Dolginov \& Mytrophanov 1976) as a promising means of driving grain alignment (Draine \& Weingartner 1996, 1997; Weingartner \& Draine 2003; Cho \& Lazarian 2005). An analytical model of alignment by RT proposed by Lazarian \& Hoang (2007) gives physical insight into the nature of the mechanism, showing that alignment occurs with long grain axes perpendicular to ${\bf B}$, i.e.\ in the expected orientation; this results not from paramagnetic relaxation, however, but from averaging of RT as the grain angular momentum precesses about ${\bf B}$.

An important objective from the observational standpoint is to attempt discrimination between these proposals by studying interstellar polarization in environments where one or other mechanism may be expected to fail. Key questions include how the efficiency of alignment varies with gas density and with the intensity and spectral energy distribution of the ambient radiation field. The polarization of starlight at visible wavelengths provides information on alignment in predominantly diffuse, translucent phases of the ISM. Observations in the near infrared allow stars behind much larger columns of dust to be explored, including those along lines of sight that intercept dense molecular clouds. Results have shown that the polarization per unit extinction declines systematically with optical depth in such regions (Goodman \etal\ 1992, 1995; Gerakines \etal\ 1995), implying that grains within clouds are generally less well aligned than those in the diffuse (intercloud) ISM. This behavior is also indicated by observations of polarized far infrared and submillimeter emission, which show evidence for reduced (but non-zero!) polarization amplitudes toward the centers of isolated globules and starless cores\footnote{This might be explained, at least in part, by the relatively low spatial resolution of the far infrared observations if changes in magnetic field direction occur in the cores on scales smaller than the beam size.} (Hildebrand \etal\ 1999; Ward-Thompson \etal\ 2000; Wolf \etal\ 2003). At first sight, this behavior does not appear to allow effective discrimination between alignment mechanisms, as it is qualitatively consistent with all of them (Lazarian \etal\ 1997). In the absence of embedded sources, alignment by radiative torques will naturally decline in efficiency within a dark cloud as the external radiation field becomes attenuated. Spin-up by Purcell torques will decline as ${\rm H \rightarrow H_2}$ conversion proceeds toward completion at higher densities, and is moreover rendered ineffective by resurfacing of grains (as evidenced by the appearance of ice mantles) which neutralizes the active centers at which H recombination preferentially occurs (Lazarian 1995)\footnote{Subsequent work by Lazarian \& Draine (1999a,\,b) showed that grains with sizes smaller than $\sim 1$\mic\ flip rapidly, averaging out Purcell's torques. This effect, termed ``thermal trapping", diminishes the effects of uncompensated torques arising from H$_{2}$ formation.}. Finally, paramagnetic or superparamagnetic alignment at thermal rotation speeds will fail as the dust and gas temperatures equilibriate at high density. Indeed, the question can be turned around: which mechanism, if any, is capable of sustaining the observed (albeit reduced) levels of grain alignment within dense molecular clouds and starless cores?

The puzzle of grain alignment within starless cores was addressed by Cho \& Lazarian (2005), who noted that RT alignment is most effective for the largest grains (mean radii $\ga 0.3$\mic) in the size distribution, which are aligned efficiently by near infrared radiation that penetrates much deeper into a cloud than ultraviolet or visible light (see also Bethell \etal\ 2007). The RT mechanism thus predicts alignment of the grains responsible for near infrared extinction and polarization at substantial optical depths within dense clouds, where other mechanisms are predicted to fail. 

Given the importance of magnetic fields and the potential of far infrared observations for mapping them in dense phases of the ISM (e.g.\ Wolf \etal\ 2003; Bethell \etal\ 2007; Pelkonen \etal\ 2007), it is timely to reexamine observational constraints placed by the polarization of starlight at shorter infrared wavelengths. The purpose of this paper is to examine in detail the correlation of alignment efficiency with extinction through representative dense clouds, with the goals of (i)~discriminating between alignment mechanisms, and (ii)~reassessing the value of near infrared polarimetry as a tracer of the internal magnetic fields. The {\it polarization efficiency}, i.e.\ the ratio of linear polarization amplitude to optical depth, $\ratlam$ (also referred to as the specific polarization), is taken as an observational measure of alignment efficiency, subject to caveats described in \S2.1 below. This approach is not new (cf.\ Gerakines \etal\ 1995) but has the advantage that the Two-Micron All-Sky Survey (2MASS) is now available to provide a coherent and essentially complete photometric database for making extinction estimates toward stars in the polarization database. We also utilize polarization data published since the earlier study, including observations of ice and silicate features as well as the near infrared continuum. 

\section{Data}
\subsection{Rationale}
The value of polarization efficiency $\ratlam$ observed at some wavelength $\lambda$ toward a background star depends on several factors in addition to the desired one, i.e.\ the efficiency with which the grains along the line of sight are aligned by the ambient magnetic field. For perfect alignment in a uniform field ${\bf B}$, $\ratlam$ changes with the viewing angle relative to ${\bf B}$ (no polarization is seen when the line of sight is parallel to the magnetic field). If field lines are twisted, or if two or more discrete clouds or clumps are present in the line of sight with different field orientations, then the net polarization is reduced relative to extinction. These unwanted effects can be minimized by careful target selection: well-sampled, isolated clouds with negligible foreground or background extinction and permeated by a uniform magnetic field with a strong transverse component are ideal. The well-known system of dark clouds in Taurus is the best available approximation: it is nearby (140\pc), lies in a direction of little general extinction, and is reasonably well-sampled by background field stars exhibiting substantial degrees of interstellar polarization (e.g.\ Whittet \etal\ 2001 and references therein). Moneti \etal\ (1984) used optical polarimetry of background stars to show that the macroscopic ${\bf B}$ field is, indeed, reasonably uniform across the cloud.

\subsection{Taurus}
Table 1 presents a compilation of broadband $K$ (2.2\mic) linear polarization and extinction data for stars subject to reddening by dust in the Taurus dark cloud. Most are background field stars that lack any direct association with the cloud; for comparison purposes, available data for a few embedded young stellar objects (YSOs) that meet our selection criteria (below) are also included in Table~1. Additional spectropolarimetric data and optical depths available for ice features (\water\ and CO) in one Taurus field star (Elias~16) are listed in Table~2. In general, polarization data were accepted only if the fractional error in the amplitude ($\Delta p/p$) is less than 25\%. Exceptions to this quality filter were made in a few cases (identified in the footnotes in Table~1) where polarimetry at shorter wavelengths (e.g.\ the $J$ and $H$ bands) could be used to corroborate the $K$ value using the usual empirical form for the wavelength-dependence of polarization (e.g.\ Whittet \etal\ 1992 and references therein). In all cases where multiple-passband polarimetry is available, a check was made to ensure that the spectral dependence is broadly consistent with interstellar polarization dominated by a single cloud. 

Infrared photometry for all selected objects was collected from the on-line 2MASS catalog and is generally accurate to 0.02~mag or better. Two methods were used to evaluate visual extinction ($\av$) and $K$-band optical depth ($\tauk$), dependent on the additional data available in each case. For objects with reliable visual photometry and spectral classifications\footnote{Spectral classifications are available from the SIMBAD database and references therein for all stars in Table~1 with Elias or HD prefixes, and also for Tamura~8.} 
(typically those with $\av\la 5$~mag), $\av$ is estimated using the relation 
\begin{equation}
\av\approx 1.1\evk
\end{equation}
where $\evk$ is the $V-K$ color excesses arising from interstellar reddening. The factor 1.1 in eq.~(1) effectively represents extrapolation of the mean interstellar extinction curve from 2.2\mic\ to the long wavelength limit (Whittet 2003), and appears to be a robust value not strongly dependent on the assumed grain model over the usual range of observed extinction curves (Whittet \& van Breda 1978). This ``$\evk$ method" (Table~1) provides $\av$ values accurate to $\pm 0.1$ or better: most of the results that use it are taken from Whittet \etal\ (2001), who provide further discussion and references to spectral classifications, visual photometry and intrinsic colors. The $K$-band optical depth then follows from the relation
\begin{equation}
\tauk = \ak/1.086 = (\av-\evk)/1.086
\end{equation}
which in combination with eq.~(1) yields $\tauk \approx 0.084\av$.

For objects in our sample lacking visual photometry, we rely on infrared photometry alone to evaluate
\begin{equation}
\av = r_1\ejk
\end{equation}
and 
\begin{equation}
\tauk = r_2\ejk
\end{equation}
(the ``$\ejk$ method"), where $r_1$ and $r_2$ are factors that depend on the form of the extinction curve over the relevant wavelengths (Martin \& Whittet 1990; Whittet \etal\ 2001). Mean values $r_1=5.3\pm 0.3$ and $r_2=0.50\pm 0.03$ determined from data for the Taurus cloud (Whittet \etal\ 2001) are adopted here (for comparison, $r_1 \approx 6.0$ and $r_2 \approx 0.54$ in the diffuse ISM). For stars with known spectral classifications, $\ejk$ is determined routinely from the observed colors to yield $\av$ values accurate to $\pm 0.4$ or better. For stars lacking spectral classifications ($\av$ values followed by colons in Table~1), $\ejk = \ejh+\ehk$ is estimated from the observed locus in the $J-H$ vs.\ $H-K$ diagram (\fig 1) by extrapolation along the appropriate reddening vector onto intrinsic color lines. This generally provides unambiguous results because background stars bright enough to be included in our sample are expected to be predominantly either late-type (K, M) giants or main-sequence stars of earlier type (typically B--G); red dwarfs distant enough to be background to the cloud are too dim at 2.2\mic\ for polarimetry to be feasible with current technology. Obviously, the absence of spectral type information reduces the reliability of $\av$ estimates ($\pm 1$~mag would be a worst case scenario) but the fractional error is nevertheless quite low for stars with the highest extinctions ($\av > 10$~mag). Note also that this technique improves upon previous estimates that adopted a standard ``one fits all" intrinsic $J-K$ color for background stars lacking spectral classifications (Tamura \etal\ 1987; Goodman \etal\ 1992; Gerakines \etal\ 1995).

The distribution of Taurus YSOs in the $J-H$, $H-K$ color-color diagram is compared with that of the field stars in \fig 1. As expected, the field stars occupy a zone consistent with normal intrinsic colors, subject to interstellar reddening. The trend for YSOs is similar, but subject to a characteristic displacement to the right that signals a contribution to their infrared colors from circumstellar matter (e.g.\ Itoh \etal\ 1996). Our estimates of $\av$ for these objects assume that the infrared excess affects predominantly the $H-K$ color and is negligible in $J-H$.

\subsection{Ophiuchus}
Although not as well sampled as Taurus for near infrared polarimetry, the \oph\ Dark Cloud provides a valid and interesting comparison. Both clouds are regions of low-mass star formation, but \oph\ is subjected to a harsher radiative environment arising from its proximity to an OB association (Sco~OB2): Liseau \etal\ (1999) deduce that the local UV field in \oph\ is typically a factor of 20--140 above the interstellar average.

Table~\ref{oph} presents a compilation of available polarization data, together with extinction estimates, for stars reddened by dust in this region, using the same selection criteria as before (\S2.2). Both background field stars and embedded YSOs are included. The color excess methods used to estimate $\av$ are the same as those described in \S2.2, except that values of the parameters $r_1=4.9\pm 0.3$ and $r_2=0.53\pm 0.05$ (eq.\,3 and 4) measured from the mean \oph\ extinction curve (Martin \& Whittet 1990; Kenyon \etal\ 1998) are adopted in place of the Taurus values. The color-excess method is unreliable for deeply embedded objects such as the Class~I YSO Oph~29 because its near infrared colors are dominated by circumstellar emission; in this case, $\av$ is estimated from the depth of the observed 9.7\mic\ silicate absorption:
\begin{equation}
\av \approx 18\tau_{9.7}
\end{equation} 
(Whittet 2003). This method assumes that the observed correlation between $\tau_{9.7}$ and $\av$ in the diffuse ISM holds inside the dense cloud. Note, however, that a recent study by Chiar \etal\ (2007) suggests a systematic difference in the sense that eq.\,5 may underestimate the true $\av$. 

\subsection{Other datasets}
Several additional sets of polarimetric data that provide useful comparisons for those discussed in detail above are listed in Table~\ref{other}. For the most part, these are self-explanatory and accessible from the single reference listed; data for individual objects are not included here. L1755 is a filamentary dark cloud cloud adjacent to and physically associated with the \oph\ cloud (Loren 1989), and L1641 is an active site of star formation in the Orion region: all point sources listed in the cited references that meet our criteria were selected. The other entries in Table~\ref{other} are grouped by spectral feature (\water-ice, CO-ice, silicate) rather than by region of sky. For a given absorption feature, the same observation provides information on the peak values of $\plam$ (from Stokes parameters $Q$ and $U$) and $\taulam$ (from intensity $I$), measured in each case relative to the continuum, leading to a measure of $\ratlam$ specific to the particles responsible for the feature (see Aitken 1996 for a review). This is straight-forward for ice features, which are only ever observed in absorption. However, silicate features may appear in absorption, emission, or a superposition of the two, and care was taken to select only those in which the observed 9.7\mic\ profile is dominated by absorption (Smith \etal\ 2000). Table~\ref{ysos} presents a comparison of polarization efficiencies for YSOs determined from ice and silicate features and $K$-band continuum data. 

The methods used to estimate $\av$ are exactly analogous to those discussed above. In the case of L1755, the \oph\ extinction parameters (\S2.3) are assumed. In the case of L1641, we adopt values based on $H-K$ colors from Casali (1995). Finally, $\av$ values calculated from the silicate optical depth use $\tau_{9.7}$ values from Smith \etal\ (2000).

\section{Results and Discussion}
\subsection{The Taurus Cloud}
The analytical tool used in this study is the plot of polarization efficiency vs.\ visual extinction ($\ratlam$ vs.\ $\av$; Figs.~2--8). For lines of sight passing through an idealized, isolated cloud consisting of denser clumps within a more diffuse envelope, $\av$ is expected to be a reasonable proxy for mean density because stars that lie behind clumps will have systematically higher extinctions. This assumption is corroborated by the fact that ices are detected only in lines of sight with $\av$ above some threshold value ($A_{\rm th}\approx 3.2$~mag in the case of the Taurus cloud; Whittet \etal\ 1988, 2001). As ices do not survive in diffuse regions regions but appear to be widespread inside molecular clouds, this extinction threshold is assumed to correspond approximately to the transition between atomic and molecular phases of the ISM (Whittet \etal\ 2001).

Plots of $\ratlam$ vs.\ $\av$ for Taurus stars are shown in \fig 2 (see Table~\ref{symbol} for a key to plotting symbols in all figures). The complete set of $K$-band data for field stars (Table~1) is included in frame~(a), and the least-squares power-law fit to this dataset (see below) is shown in all frames (thick diagonal line). The thinner (parallel) diagonal lines, which lie 0.34~dex above and below the best fit, represent the approximate upper and lower bounds of the distribution. Frames~(b)--(d) in \fig 2 show subsets of the data in frame~(a) according to the method used to determine $\av$ and $\tauk$ ($\evk$ method; $\ejk$ method without or with spectral types, respectively). The $\av$ range spanned by these subsets is different, as expected, but they nevertheless show reasonable consistency with the common fit, confirming that no major systematic errors exist between them. It is also clear that there is real scatter in the data: the degree of scatter is significantly greater than that implied by observational error, at least for the stars with spectral classifications. This is to be expected for reasons explained in \S2.1.

\fig 2a shows that a power law provides a fair representation of the correlation between $\ratk$ and $\av$ for field stars, in agreement with the results of Gerakines \etal\ (1995). The fit is of the form
\begin{equation}
\ratk = a (\av)^{-b}
\end{equation}
where $a=5.3\pm 0.3$ and $b=0.52\pm 0.07$. Power-law behavior is consistent with the dependence of magnetic field strength on density for classical DG alignment (Vrba \etal\ 1981), although the applicability of such a model is questionable (see Gerakines \etal\ 1995 for discussion and caveats). We use it here merely as an empirical law that can be used to predict (by extrapolation) the alignment to higher extinctions.

\fig 2e compares the distribution of 2.2\mic\ broadband data for Taurus YSOs (Table~1) with the fit to field stars. A systematic difference would be expected if, for example, radiation from embedded stars influences grain alignment in surrounding regions of the cloud. There is, indeed, some evidence for higher polarization efficiency toward the YSOs (6 out of 7 lie above the correlation line) but the sample is too small to make a strong statement.

\fig 2f plots two $\ratlam$ values based on ice-feature spectropolarimetry of Elias~16 (Table~2). Polarization is measured at the peak of each feature relative to the continuum, and ratioed with the corresponding optical depth (Hough \etal\ 1988, 2008). The results are consistent with the $K$-band continuum datum for this star (Table~2) and with the general trend implied by the fit to all field star data in \fig 2. This is an important result as it implies that ice-mantled grains in the line of sight are aligned to a similar degree compared with those responsible for the continuum polarization, even though the latter must include a contribution from unmantled grains. The detection of polarization in the CO ice feature is an especially significant result (Hough \etal\ 2008; Crysostomou \etal\ 1996), as the low sublimation temperature of CO ($\sim 17$~K in pure form) implies that it can exist in solid phase only in the coldest, densest regions along the line of sight.

The vertical dotted line in each frame of \fig 2 denotes the threshold visual extinction ($A_{\rm th}= 3.2\pm 0.1$~mag) for detection of \water-ice in the Taurus cloud, as discussed above. Below this threshold the grains resemble those in the diffuse ISM, lacking a volatile component; above it they acquire icy mantles, as evidenced by the onset of the 3.0\mic\ \water-ice absorption feature in the spectra of background stars. CO$_2$ and CO ices are also observed with somewhat higher thresholds ($A_{\rm th}\sim 4.3$ and 6.7~mag, respectively; Whittet \etal\ 2007; Chiar \etal\ 1995). A key question is whether the ability of a grain to align in the ambient magnetic field undergoes any substantive change with the growth of a mantle. The distribution of data (e.g.\ \fig 3) shows no evidence for a significant discontinuity or change in slope near the threshold value of $\av$. This general result, together with the presence of polarized ice features in at least one line of sight through the cloud, clearly demonstrates that grains are, indeed, capable of being aligned in environments where they are subject to ice-mantle growth. As the accretion process that leads to mantle formation is expected to inactivate the sites on a grain surface that drive Purcell alignment by H recombination (\S1), this finding constitutes further evidence against the Purcell mechanism as the dominant means of alignment for extinctions $\av \ga A_{\rm th}$.

The general trend of declining polarization efficiency with extinction led some investigators to propose that grains within dense clumps make no contribution to the net observed polarization of background stars (Goodman \etal\ 1992, 1995; Arce \etal\ 1998). Observations of ice-feature polarization provide counter-examples but it is nevertheless useful to consider this scenario as a limiting case, which we refer to as the ``skin-depth" model. The structure of the Taurus cloud system resembles a diffuse screen with embedded clumps, as described above, such that the maximum opacity of the screen is approximately equivalent to the ice threshold extinction (see, e.g., \fig 1 of Whittet \etal\ 2004). Suppose that {\it all\/} of the observed $K$-band polarization arises in the diffuse screen at extinctions in the range $0< \av < A_{\rm th}$ ($A_{\rm th}=3.2$~mag). Then, for lines of sight that intersect clumps ($\av > A_{\rm th}$) $p_K$ is independent of $\av$, and thus $\ratk\propto [\av]^{-1}$ (assuming $\tauk\propto\av$). This form of behavior is compared with the data in \fig 3, the constant of proportionality being set by the mean of all data below the ice extinction threshold. A power-law index of $-1$ obviously represents a much steeper decline than our best fit ($-0.52\pm 0.07$). The skin-depth model is nevertheless reasonably consistent with the data for $\av\la 10$~mag (\fig 3), failing at higher extinction: at $\av\sim 25$~mag, for example, it predicts $\ratk$ to be a factor of $\sim3$--4 less than observed.

The ``Taurus trend" represented by the best fit and upper and lower bounds in \fig 2 is used as a template for comparison with the other samples discussed in the following sections.

\subsection{The Ophiuchus Cloud}
Polarization efficiency vs.\ visual extinction is plotted for \oph\ stars in \fig 4. Broadband $K$ data are from Table~3, and spectropolarimetric data for YSO Oph~29 (\water-ice and silicate) are from references listed in Table~4. The diagonal lines show the fit and upper/lower bounds to Taurus stars, as in \fig 2. A fit to \oph\ field stars (not shown) yields a power-law index $-0.45\pm 0.10$, not significantly different from the Taurus result. In general, both field stars and YSOs in \oph\ are consistent with the trend implied by the Taurus data. No systematic difference is apparent between \oph\ field stars and YSOs in the $\av$-range of overlap ($\sim 5$--15~mag) in \fig 4. At higher extinction ($\av\approx 15$--30~mag), there is evidence for greater polarization efficiency in the YSO data. Interestingly, the \water-ice datum (triangle) for Oph~29 is significantly lower (and consistent with the Taurus line) compared with the silicate and $K$-continuum data in the same line of sight (see also Table~\ref{ysos}).

\subsection{L1755}
Polarization efficiency vs.\ visual extinction for L1755 is plotted in \fig 5. All points represent field stars behind this filamentary dark cloud (Goodman \etal\ 1995). The diagonal lines show the fit and upper/lower bounds of the Taurus data, as before. The decline in polarization efficiency with increasing extinction appears to be significantly steeper in L1755 compared with Taurus (or Ophiuchus), and is more consistent with the skin-depth scenario described in \S3.1 above, illustrated by the dotted line in \fig 5. This finding is in agreement with the results of Goodman \etal\ (1995).

It is not clear why L1755 should display this form of behavior, which is unique amongst the regions we investigated. It might be that physical conditions leading to grain alignment are lacking inside this cloud, that the internal grains are effectively spherical, or that peculiarities in magnetic field structure and/or viewing geometry conspire to systematically reduce the observed $\ratlam$. It should also be noted that the total range in $\av$ covered by the observations (2--8~mag) is small compared with that of the other clouds studied: it would be of obvious interest to obtain polarization data for stars with $\av>10$~mag toward L1755 to see if the trend continues.

\subsection{L1641}
L1641 is a turbulent region of clustered low-mass star formation (e.g.\ Stanke \& Williams 2007), and as such it differs markedly from the prototypical dark cloud. Sources with available polarimetry (Casali 1995) are embedded YSOs originally detected by IRAS (Strom \etal\ 1989). The level of polarization efficiency observed in L1641 is nevertheless remarkably consistent with that in our template cloud, as illustrated in \fig 6. All 15 objects in the sample lie within the bounds implied by the Taurus data. 
 
\subsection{Feature polarization in YSOs}
Young stellar objects with feature polarization data available are listed in Table~\ref{ysos}. Out of 19 in the set, all but two (Oph~29 and SVS~13) are considered to be high-mass objects that seem likely to exert a strong influence on physical conditions in their vicinity. \fig 7 plots feature polarization data for ices and silicates in frames~(a) and (b), respectively, together with available $K$-band data for the same objects in frame~(c). The plot for ices includes one datum for CO, the other points being for \water. In each frame, the locus of each individual point may be described as either ``normal" or ``enhanced" in terms of polarization efficiency with respect to the Taurus template distribution, ``enhanced" meaning significantly above the upper bound. YSOs with enhanced polarization efficiencies according to this definition are identified by values in bold type in Table~\ref{ysos}. Some objects, notably Mon~R2 no.\,2 and OMC--1~BN, show similar behavior at all wavelengths. Others show discrepancies that seem likely to reflect differing physical conditions in the individual lines of sight, perhaps leading to differences in the distributions of volatile and non-volatile grain material as well as in the efficiency of alignment. 

The ice features are more likely to reflect conditions within the molecular clouds, whereas the silicate features are more likely to include a circumstellar component. The fact that enhancements are observed for ices in 6 out of 9 lines of sight toward YSOs (and 5 out of 7 toward high-mass YSOs) therefore seems highly significant: this result implies that grain alignment is more efficient in molecular clouds that contain regions of active star formation (especially {\it massive\/} active star formation), compared with those that do not. Independent support for this conclusion is provided by observations of continuum circular polarization, the highest amplitudes of which are likewise found in lines of sight to massive YSOs (e.g.\ Buscherm\"ohle \etal\ 2005, Clayton \etal\ 2005, and references therein), and are best explained by models that assume efficient grain alignment in the envelope of the YSO (Whitney \& Wolff 2002).

\subsection{Overview}
\fig 8 displays summary plots that include data for all regions, passbands and features in this study, with field stars and YSOs plotted in frames~(a) and (b), respectively. The field-star data (\fig 8a) are consistent with a power-law dependence for polarization efficiency on extinction in the observed range $0<\av<25$, with scatter of approximately a factor~2 about the mean relationship for the template cloud. Results for field stars provide unequivocal evidence that ice-mantled grains are aligned inside dark clouds, and that the onset of mantle growth has no dramatic effect on net alignment.

The YSO data (\fig 8b) are distributed either close to or significantly above the mean power-law relationship for field stars behind the template cloud. YSOs typically lie above field stars at comparable extinctions within the range of overlap ($1<\av<25$), and enhancements in $\ratlam$ by factors of up to $\sim 6$ with respect to the power-law fit occur toward some YSOs (predominantly those of high mass and with extinction $\av>20$~mag). Enhancements of this order are seen in the ice features as well as in the silicate feature and $K$-band continuum. The simplest explanation for the distribution of YSO data is that each line of sight contains a ``quiescent cloud" component that follows the general trend, and a ``modified" component in which alignment is enhanced by local conditions, the relative contributions of each differing from one line of sight to another. It is logical to assume that alignment in the modified component may be driven by radiation from the embedded stars themselves.

\section{Modeling}
We noted in \S1 that a systematic decline in alignment efficiency with increasing density inside cold (starless) molecular clouds is a prediction of all the major alignment mechanisms (Lazarian \etal\ 1997). In regions cold enough for solid CO to accumulate on the grains (Chrysostomou \etal\ 1996) we expect catastrophic failure of paramagnetic or superparamagnetic DG alignment at thermal rotation speeds. Significant alignment at suprathermal rotation speeds by the Purcell mechanism also appears to be ruled out (\S1); however, alignment by radiative torques seems more promising as a mechanism that can operate within molecular clouds, albeit with reduced efficiency as the external radiation field becomes progressively attenuated with increasing $\av$. Extinction by dust naturally changes the spectrum as well as the intensity of the ambient radiation field, effectively blocking the shorter wavelengths whilst allowing the infrared to penetrate much further into a cloud. A key point that arises from theory (Cho \& Lazarian 2005; Lazarian \& Hoang 2007) is that the efficiency of the RT mechanism increases systematically with grain size and is optimized for grains with dimensions comparable with the wavelength. Thus, the largest grains in the size distribution may be aligned by near infrared radiation in regions of a cloud that are optically thick to shorter wavelengths.

The alignment of grains by radiative torques (Dolginov \& Mytrophanov 1976; Draine \& Weingartner 1996, 1997; Lazarian \& Hoang 2007) is used to calculate the expected relationship between polarization efficiency and visual extinction through an idealized homogeneous cloud. We assume the cloud to have gas density $n_{\rm H}=10^{4}~{\rm cm^{-3}}$, gas and dust temperatures $T_{\rm gas}=20$~K and $T_{\rm dust}=15$~K, radius $r_{\rm cloud}=0.5$\,pc and extinction $\av=12$~mag (center to edge). Oblate spheroidal grains with axial ratio $a/b = 0.5$ and sizes following the MRN distribution (Mathis, Rumple \& Nordseick 1977) are adopted with limiting sizes $a_{\rm min}=0.005$\mic\ and $a_{\rm max}=0.3$\mic. In calculating radiative torques, we take into account that real grains are irregular and can have significant helicity. Although crucial to RT, however, irregularity and helicity are higher order effects when we deal with grain extinction. Indeed, it is customary to approximate irregular but swiftly rotating grains by spheroids, and in our simulations of light
propagation we follow this custom.

Radiative torque efficiencies were first calculated for irregular grains using the publicly available code DDSCAT (Draine \& Flatau 1994) for the entire spectrum of the mean interstellar radiation field (0.1--100\mic; Mathis, Mezger \& Panagia 1983). Grains of shape~1 in Draine \& Weingartner (1996, 1997) were adopted, each consisting of 13 identical cubes, and the calculations were carried out over the full range of grain sizes. We then calculated the angular velocity of grains spun up by radiative torques arising from the attenuated radiation field for grain size in the range $a_{\rm min}$ to $a_{\rm max}$, assuming perfect coupling of the grain axis of major inertia and angular momentum, and with both axes parallel to the radiation beam. Other alignment mechanisms such as DG and Purcell alignment (\S1) were disregarded. The minimum critical size for alignment, $a_{\rm crit}$ was then deduced for the criterion of suprathermal rotation, $\omega_{\rm rad}/\omega_{\rm therm}\ge 3$. This criterion is based on calculations in Hoang \& Lazarian (2007), who studied the influence of gas bombardment on alignment by radiative torques using the Langevin equation approach: it was shown that when grains are aligned by radiative torques at an attractor point with an angular momentum greater than a value about $3 J_{\rm therm}$, they are not randomized by gas  bombardment. The polarization arising from aligned grains is given by 
\begin{equation}
p_{\lambda}=\int ds \int_{a_{\rm crit}}^{a_{\rm max}} \frac{1}{2}\,R C a^{-3.5}\pi a^{2} Q_{\rm pol}(a,\lambda)\,da
\end{equation}
where $C$ is a normalization constant of the size distribution, $R$ is the Rayleigh reduction factor that characterizes alignment (Roberge \& Lazarian 1999), $Q_{\rm pol}=Q_{\perp}-Q_{\|}$ is the polarization efficiency, and $Q_{\perp}, Q_{\|}$ are extinction efficiencies for the axis of major inertia perpendicular and parallel to the electric field, respectively. The corresponding optical depth is given by
\begin{equation}
\tau_{\lambda}=\int ds \int_{a_{\rm min}}^{a_{\rm max}} C a^{-3.5}\pi a^{2} Q_{\rm ext}(a,\lambda) \,da
\end{equation}
where $Q_{\rm ext}=(2Q_{\perp}+Q_{\|})/3$ is the total extinction. The polarization specified by eq.\,(7) is optimized when the magnetic field vector ${\bf B}$ is transverse to the line of sight, but in general the direction of ${\bf B}$ is arbitrary; this may be taken into account by replacing $R$ in eq.\,(7) with $R^{*}=R\,{\cos}^{2}\zeta$, where $\zeta$ is the angle between ${\bf B}$ and the plane of the sky (see e.g.\ \fig 3 of Roberge \& Lazarian 1999). $R^{*}$ is thus a composite quantity dependent on the direction of ${\bf B}$ as well as on the degree of grain alignment characterized by the Rayleigh reduction factor.

Systematic changes in $a_{\rm crit}$ as a function of the radiation field leads to a prediction of the wavelength of maximum polarization ($\lmax$) with respect to extinction: this prediction is shown and compared with observational data for the Taurus cloud in \fig 9. As the interstellar polarization curve typically peaks at visible wavelengths, empirical determination of $\lmax$ (from the Serkowski formula) requires availability of polarimetric data in the visible as well as the near infrared, and is thus possible only for lines of sight with relatively low extinction ($\av \la 6$~mag with one exception; Whittet \etal\ 2001). Considerable scatter is evident in \fig 9, but the model reproduces the general trend in the data toward longer $\lmax$ at larger $\av$. Note that no changes in the size distribution with $\av$ are assumed in the model: variations in $\lmax$ are a consequence of variations in the average size of the {\it aligned\/} component of the grains.

The dependence of polarization efficiency on extinction predicted by the model is shown for several values of $R^{*}$ and compared with observational data for the Taurus cloud in \fig 10. In addition to the near infrared ratio $\ratk$ (\fig 10a) we also plot $\pmax/\av$ (\fig 10b). The advantage of the latter quantity is that the formal errors in the observational data are smaller (because $\pmax$ is determined from a fit to several passbands instead of from just one) at the cost of a much smaller sample. Our calculations illustrate that the RT model is capable of producing alignment sufficient to explain the observed polarization efficiencies in the Taurus cloud for extinctions up to at least $\av=10$~mag. Indeed, the predicted rate of decline in $p/\tau$ with $\av$ is less steep than the observed trend, especially in the near infrared (\fig 10a). This may easily be explained by physical differences in the real cloud compared with the idealized model cloud, such as twists in magnetic field direction and superposition of discrete cloud components with different field orientations (\S2.1).

In our simple model, the alignment mechanism becomes inoperative as $\av$ approaches 10~mag because even the largest grains fail to align ($a_{\rm crit}\rightarrow a_{\rm max}$). However, RT alignment is still possible at higher extinctions if the grains are subject to growth (leading to an increase in $a_{\rm max}$). The grains are, indeed, expected to grow inside the cloud by processes of both mantle accretion and coagulation (e.g.\ Draine 1985; Wurm \& Schnaiter 2002). It is important to realize, also, that $\av$ is not necessarily a reliable measure of radiation field, as it accounts for attenuation only in one direction: inhomogeneities in the cloud may lead to deeper penetration of the external field in a given line of sight than expected for the nominal value of $\av$ (see Andersson \& Potter 2007 for detailed discussion). 

We noted in \S2.3 that the \oph\ cloud (\fig 4) is subject to a much stronger ambient radiation field compared with Taurus, arising from the proximity of OB stars, yet polarization efficiencies are remarkably consistent between the two clouds at similar values of $\av$. It is possible, of course, that differences in (e.g.) cloud structure or magnetic field geometry (\S2.1) conspire to reduce $\ratlam$ in our line of sight to \oph, and thus negate any radiation-driven enhancement in alignment relative to Taurus. Moreover, the radiation field near \oph\ is enhanced primarily in the UV and may not be so different at the near infrared wavelengths that drive alignment of the large grains responsible for near infrared polarization in our model. For the largest grains, RT alignment is predicted to approach saturation in moderate radiation fields (Lazarian \& Hoang 2007), at which point further increases in intensity will have no further effect on alignment.

\section{Conclusion and a suggestion for future research}
This paper yields the following primary results:

\begin{enumerate}
\item Within the limits of our sample, the general trend of declining polarization efficiency with optical depth in quiescent regions of dense clouds is well described by a power law of the form $\ratlam \propto [\av]^{-b}$, with $b\approx 0.52$.
\item Polarization efficiencies observed toward YSOs indicate that grain alignment tends to be enhanced in molecular clouds that contain regions of active massive star formation.
\item Although our results are based on a dataset dominated by broadband polarimetry, observations of ice-feature polarizations in a small subset of the targets are particularly powerful in elucidating the efficiency of grain alignment in molecular clouds.
\item Of extant grain alignment models, the radiative torques mechanism seems most effective in producing degrees of alignment adequate to explain the observations.
\end{enumerate}

We note in conclusion that spectropolarimetry of ice features may provide not only the most stringent method of testing alignment mechanisms but also the most promising technique for mapping magnetic fields internal to dense molecular clouds. It was pointed out some years ago that the observed reduction in polarization efficiency with extinction compromises the value of near infrared broadband polarimetry for mapping (e.g.\ Goodman \etal\ 1992, 1995). This is because the net polarization in a given line of sight tends to be dominated by dust in the diffuse outer layers of a cloud, where alignment is most efficient (the skin-depth scenario; \S3.1): there is no way to separate the effect of the ``skin" to reveal the internal field because all aligned grains in the line of sight contribute to the observed continuum polarization. However, the excess polarization observed in an ice absorption feature naturally allows this separation, because only grains internal to the cloud (where temperatures are low enough for ice mantles to from) will contribute to the excess. The amplitude and position angle of the feature component of the polarization thus traces the internal field. To date, the number of relevant observations is extremely small, dominated by bright, massive YSOs (cf.\ Table~\ref{ysos}), and includes only a single field star (Elias~16; Table~2). It should be possible in the future to carry out narrow-band imaging polarimetry that allows ice-feature polarizations and optical depths to be measured simultaneously toward large numbers of faint background stars over extended regions of molecular clouds.

\acknowledgments
This research utilized data products from the Two Micron All Sky Survey, a joint project of the University of Massachusetts and the Infrared Processing and Analysis Center, funded by the National Aeronautics and Space Administration (NASA) and the National Science Foundation (NSF). Extensive use was also made of the SIMBAD database, operated at CDS, Strasbourg, France. A.L.\ and T.H.\ acknowledge the support of the NSF Center for Magnetic Self-Organization in Laboratory and Astrophysical Plasmas and of NSF grant AST~0507164.

\clearpage

\begin{deluxetable}{lllcccc} 
\tabletypesize{\scriptsize} 
\tablecaption{Broadband polarization and extinction data for Taurus stars. \label{tauk}} 
\tablewidth{0pt} 
\tablehead{\colhead{Star$^a$} & \colhead{$p_K$} & \colhead{$\Delta p_K$} & 
\colhead{Ref.$^b$} & \colhead{$A_V$} & \colhead{Method} & \colhead{$p_K/\tau_K$}\\
& (\%) &&& (mag)}
\startdata
\underline{Field stars:}\\
\\
Elias 3		& 0.65	& 0.02	& 1, 2 & 10.0	& $\ejk$	& 0.8 \\
Elias 6		& 0.90	& 0.08	& 3    & 6.8	& $\ejk$	& 1.6 \\
Elias 13	& 0.85	& 0.10	& 3    & 11.7	& $\ejk$	& 0.9 \\
Elias 15	& 2.08	& 0.20	& 3, 4 & 15.3	& $\ejk$	& 1.6 \\
Elias 16	& 2.47	& 0.08	& 4, 5 & 24.1	& $\ejk$	& 1.2 \\
Elias 19	& 1.20	& 0.03	& 1	   & 2.5	& $\ejk$	& 5.7 \\
Elias 29	& 0.94	& 0.09	& 1    & 4.2	& $\ejk$	& 2.7 \\
Elias 30	& 1.14	& 0.04	& 2    & 4.9	& $\ejk$	& 2.8 \\
GJL 41919.7+263926~~&0.76	& 0.07	& 6    & 5.5:	& $\ejk$	& 1.6 \\
GJL 42003.6+264232	& 0.38	& 0.06	& 6    & 3.3:	& $\ejk$	& 1.4 \\
GJL 42014.0+263419	& 0.91	& 0.13	& 6    & 11.6:	& $\ejk$	& 0.9 \\
GJL 42041.6+263553	& 0.87	& 0.12	& 6    & 8.8 	& $\ejk$	& 1.2 \\
GJL 42046.0+263313	& 0.80	& 0.16	& 6    & 6.7: 	& $\ejk$	& 1.4 \\
GJL 42047.6+264121	& 0.51	& 0.06	& 6    & 3.3: 	& $\ejk$	& 1.8 \\
GJL 42052.4+262944	& 0.98	& 0.12	& 6    & 5.5: 	& $\ejk$	& 2.1 \\
GJL 42055.3+264311	& 0.41	& 0.05	& 6    & 2.1:	& $\ejk$	& 2.3 \\
GJL 42124.9+263323	& 0.65	& 0.12	& 6    & 4.3:	& $\ejk$	& 1.8 \\
GJL 42127.8+262153	& 1.29	& 0.09	& 6    & 4.2:	& $\ejk$	& 3.7 \\
GJL 42129.3+262645	& 0.65	& 0.08	& 6    & 3.0:	& $\ejk$	& 2.6 \\
GJL 42134.8+262000	& 1.33	& 0.11	& 6    & 4.0:	& $\ejk$	& 4.0 \\
GJL 42146.3+262205	& 0.99	& 0.05	& 6    & 3.8:	& $\ejk$	& 3.1 \\
GJL 42211.6+262225	& 1.01	& 0.05	& 6    & 4.0: 	& $\ejk$	& 3.0 \\
GJL 42232.2+261258	& 0.65	& 0.12	& 6    & 2.7:	& $\ejk$	& 2.9 \\
GJL 42234.1+262246	& 1.47	& 0.05	& 6    & 4.9:	& $\ejk$	& 3.6 \\
HD 28170	& 0.38	& 0.07	& 1, 7 & 1.32	& $\evk$	& 3.4 \\
HD 28225	& 0.34	& 0.02	& 2    & 1.22	& $\evk$	& 3.3 \\
HD 28975	& 0.82	& 0.11	& 1    & 1.75	& $\evk$	& 5.6 \\
HD 29333	& 1.03	& 0.07	& 1    & 1.98	& $\evk$	& 6.2 \\
HD 29647	& 0.64	& 0.06	& 1    & 3.63	& $\evk$	& 2.1 \\
HD 29835	& 0.80	& 0.06	& 1    & 1.14	& $\evk$	& 8.4 \\
HD 30168	& 0.59	& 0.07	& 1    & 1.18	& $\evk$	& 6.0 \\
HD 30675	& 0.64	& 0.11	& 1    & 1.54	& $\evk$	& 5.0 \\
HD 279652	& 0.20	& 0.04	& 1, 7 & 0.99	& $\evk$	& 2.4 \\
HD 279658	& 0.38	& 0.05	& 1, 7 & 1.87	& $\evk$	& 2.4 \\
HD 283367	& 0.36	& 0.04	& 2	   & 1.95	& $\evk$	& 2.2 \\
HD 283637	& 0.54	& 0.05	& 1    & 2.28	& $\evk$	& 2.8 \\
HD 283642	& 0.42	& 0.03	& 2    & 2.19	& $\evk$	& 2.3 \\
HD 283643	& 0.29	& 0.06	& 2, 7 & 1.66	& $\evk$	& 2.1 \\
HD 283701	& 0.73	& 0.07	& 1    & 2.53	& $\evk$	& 3.4 \\
HD 283725	& 1.12	& 0.08	& 1    & 1.75	& $\evk$	& 7.6 \\
HD 283757	& 0.67	& 0.09	& 2    & 1.70	& $\evk$	& 4.7 \\
HD 283800	& 0.69	& 0.07	& 1    & 1.64	& $\evk$	& 5.0 \\
HD 283809	& 1.48	& 0.09	& 1    & 5.70	& $\evk$	& 3.1 \\
HD 283812	& 1.11	& 0.06	& 1    & 1.91	& $\evk$	& 6.9 \\
HD 283815	& 0.57	& 0.05	& 2    & 1.91	& $\evk$	& 3.6 \\
HD 283877	& 0.37	& 0.08	& 2, 7 & 0.72	& $\evk$	& 6.1 \\
HD 283879	& 0.86	& 0.05	& 2    & 3.33 	& $\evk$	& 3.1 \\
TNS 8		& 2.7	& 0.3 	& 4    & 21.5 	& $\ejk$	& 1.5 \\
TNS 12		& 1.5	& 0.2 	& 4    & 6.9: 	& $\ejk$	& 2.6 \\
TNS 13		& 0.8	& 0.2	& 4    & 2.0:	& $\ejk$	& 4.8 \\
TNS 15		& 1.3	& 0.3	& 4    & 6.2:	& $\ejk$	& 2.5 \\
TNS 16		& 2.0	& 0.2 	& 4    & 5.3:	& $\ejk$	& 4.5 \\
TNS 17		& 1.2	& 0.1	& 4    & 4.0: 	& $\ejk$	& 3.6 \\
\\
\underline{Embedded YSOs:} \\
\\
Elias 1 (V892 Tau)		& 1.83 & 0.10 & 1, 3 & 13.8 & $\ejk$ & 1.6 \\		
Elias 5 (DG Tau)		& 2.10 & 0.12 &  3   & 5.2: & $\ejk$ & 4.8 \\				
Elias 12 (V807 Tau)		& 0.90 & 0.18 &  3   & 1.5  & $\ejk$ & 7.2 \\
Elias 18 (IC 2087 IR)	& 1.48 & 0.10 &  3   & 22.2 & $\ejk$ & 0.8 \\
GJL 41857.5+265029		& 1.54 & 0.11 &	 6   & 8.4: & $\ejk$ & 2.2 \\ 
GJL 41943.6+263854$^c$	& 1.39 & 0.09 &  6   & 7.5	& $\ejk$ & 2.2 \\
HN Tau					& 2.74 & 0.57 &  3   & 6.5: & $\ejk$ & 5.0 \\
\\
\enddata 
\tablenotetext{a}{Elias, GJL, and TNS identifications are from Elias 1978b, Goodman \etal\ 1992, and Tamura \etal\ 1987, respectively; the Goodman \etal\ numbering is based on 1950 coordinates.}
\tablenotetext{b}{Sources of polarimetry: (1)~Whittet \etal\ 1992 with adjustment to the standard $K$ passband (see Gerakines \etal\ 1995); (2)~Whittet \etal\ 2001; (3)~Tamura \& Sato 1989; (4)~Tamura \etal\ 1987; (5)~Hough \etal\ 1988; (6)~Goodman \etal\ 1992; (7)~Extrapolation based on fit to $p(\lambda)$.}
\tablenotetext{c}{IRAS 04196+2638; YSO status determined by Luhman \etal\ 2006.}
\end{deluxetable}

\begin{deluxetable}{lccccl} 
\tabletypesize{\scriptsize} 
\tablecaption{Comparison of broadband and feature data for Taurus field star Elias~16. \label{tice}} 
\tablewidth{0pt} 
\tablehead{\colhead{Description} & \colhead{$\lambda$} & \colhead{$p_\lambda$} & \colhead{$\tau_\lambda$} 
& \colhead{$p_\lambda/\tau_\lambda$} & \colhead{Ref.$^a$} \\ & ($\mu$m)}
\startdata 
$K$ band            & 2.2  & $2.47\pm 0.08$ & $2.02\pm 0.06$ & $1.22\pm 0.08$ & 4, 5, 8 \\
\water\ ice feature & 3.0  & $1.26\pm 0.10$ & $1.43\pm 0.15$ & $0.88\pm 0.16$ & 5, 9 \\
CO ice feature      & 4.67 & $1.20\pm 0.10$ & $1.06\pm 0.02$ & $1.13\pm 0.12$ & 10 \\
\enddata 
\tablenotetext{a}{References: (4)--(5) as Table~1; (8)~$K$-band optical depth from the present work; 
(9)~\water-ice optical depth averaged from Whittet \etal\ 1988 and Smith \etal\ 1993; (10)~Hough \etal\ 2008.}
\end{deluxetable}

\clearpage

\begin{deluxetable}{llccccc} 
\tabletypesize{\scriptsize} 
\tablecaption{Broadband polarization and extinction data for \oph\ stars. \label{oph}} 
\tablewidth{0pt} 
\tablehead{\colhead{Star$^a$} & \colhead{$p_K$} & \colhead{$\Delta p_K$} & 
\colhead{Ref.$^b$} & \colhead{$A_V$} & \colhead{Method} & \colhead{$p_K/\tau_K$}\\
& (\%) &&& (mag)}
\startdata 
\underline{Field stars:}\\
\\
HD 145502 ($\nu$ Sco) 	& 0.26 & 0.01 & 13 		&  1.06 & $\evk$ & 2.9 \\
HD 147084 ($o$ Sco)		& 0.91 & 0.04 & 12 		&  2.70 & $\evk$ & 4.0 \\
HD 147283				& 0.41 & 0.04 & 1		&  2.50 & $\evk$ & 2.0 \\
HD 147888 (\oph~D)		& 0.71 & 0.05 & 13		&  1.85 & $\evk$ & 4.6 \\
HD 147889				& 1.12 & 0.02 & 11, 12	&  4.60 & $\evk$ & 2.9 \\
HD 147932 (\oph~C)		& 0.87 & 0.05 & 13		&  2.10 & $\evk$ & 4.9 \\
HD 147933/4 (\oph~A/B)	& 0.58 & 0.03 & 13		&  1.79 & $\evk$ & 3.9 \\
HD 149757 ($\zeta$ Oph)	& 0.29 & 0.02 & 12 		&  1.06 & $\evk$ & 3.3 \\
Oph 7 (S3)				& 1.30 & 0.06 & 11		&  7.9  & $\ejk$ & 1.5 \\
Oph 8 (S4)				& 1.88 & 0.05 & 11		&  8.5  & $\ejk$ & 2.0 \\
Oph 11 (GSS15)			& 1.04 & 0.04 & 11		&  9.7  & $\ejk$ & 1.0 \\
Oph 16 (SR3)			& 1.40 & 0.08 & 11, 12	&  6.5  & $\ejk$ & 2.0 \\
Oph 25 (S1)				& 2.17 & 0.04 & 11, 12	& 13.1  & $\ejk$ & 1.5 \\
Oph 35 (VSSG13)			& 1.65 & 0.04 & 11		& 11.6  & $\ejk$ & 1.3 \\
\\
\underline{Embedded YSOs:} \\
\\
GSS 26					& 5.3  & 0.4  & 11		&  25:  & $\ejk$ & 2.0 \\
HD 150193 (V2307 Oph)	& 1.44 & 0.03 & 1		&  5.9  & $\ejk$ & 2.2 \\
Oph 14 (V2246 Oph)		& 1.85 & 0.05 & 14		&  9.8  & $\ejk$ & 1.8 \\
Oph 20 (VSSG1)			& 2.33 & 0.12 & 11		&  18:  & $\ejk$ & 1.2 \\
Oph 22 (DoAr 24E)		& 0.85 & 0.04 & 11		&  9.2  & $\ejk$ & 0.9 \\
Oph 23 (S2)				& 3.1  & 0.2  & 11		&  14:  & $\ejk$ & 2.0 \\
Oph 24 (IRAS 16233$-$2409)&2.04& 0.09 & 11		& 11.9  & $\ejk$ & 1.6 \\
Oph 27 (GSS39)			& 3.6  & 0.3  & 11		&  17:  & $\ejk$ & 2.0 \\
Oph 29					& 6.69 & 0.10 & 15		&  30:  & $\tau$(sil) & 2.7 \\
Oph 33 (VSSG17)			& 4.2  & 0.3  & 11		&  26:  & $\ejk$ & 1.5 \\
Oph 36 (VSSG14)			& 2.02 & 0.14 & 11		&  8.5: & $\ejk$ & 2.2 \\
\enddata 
\tablenotetext{a}{Oph designations from Elias 1978a; GSS designations from Grasdalen, Strom \& Strom 1973; VSSG designations from Vrba \etal\ 1975.}
\tablenotetext{b}{References: (1) as Table~1; (11) Wilking \etal\ 1979; (12) Wilking \etal\ 1980; (13) Wilking \etal\ 1982; (14) Martin \etal\ 1992; (15) Hough \etal\ 1989.}
\end{deluxetable}

\clearpage

\begin{deluxetable}{lcccl}
\tabletypesize{\scriptsize} 
\tablecaption{Summary of comparison data. \label{other}} 
\tablewidth{0pt} 
\tablehead{\colhead{Region/sample} & \colhead{$N_{\rm stars}$} & \colhead{$\lambda$} 
& \colhead{$\av$} & \colhead{Reference} \\ && ($\mu$m) & method}
\startdata 
L1755 field stars 				& 22 & 2.2 & $\ejk$ 	 & Goodman et al.\ 1995 \\
L1641 YSOs  					& 15 & 2.2 & $\ehk$ 	 & Casali 1995$^a$ \\
\water-ice features in YSOs 	&  8 & 3.0 & $\tau$(sil) & Holloway et al.\ 2002$^b$ \\
CO-ice feature in YSO W33A		&  1 & 4.7 & $\tau$(sil) & Chrysostomou et al.\ 1996 \\
Silicate features in YSOs 		& 19 & 9.7 & $\tau$(sil) & Smith et al.\ 2000$^c$ \\
\enddata 
\tablenotetext{a}{Only point-like sources with $p_K<p_H$ were selected.}
\tablenotetext{b}{Cited reference includes a compilation of data from earlier literature.}
\tablenotetext{c}{Only sources with silicate spectra dominated by absorption were selected (see Table~\ref{ysos}).}
\end{deluxetable}


\begin{deluxetable}{lcccc} 
\tabletypesize{\scriptsize} 
\tablecaption{Broadband and feature polarization efficiencies compared for YSOs.$^a$ \label{ysos}} 
\tablewidth{0pt} 
\tablehead{\colhead{YSO} & \colhead{$\av$} & \colhead{$\ratk$} & \colhead{$\ratlam$$^{\,b}$} & \colhead{$\ratlam$} \\ 
	& \colhead{(mag)} & & \colhead{ices~} & \colhead{silicate}}
\startdata 
AFGL 490		& 34 &    2.0    & {\bf 5.1} &    2.2    \\
AFGL 896		& 68 &    ---    & {\bf 3.5} &    1.6    \\
AFGL 961		& 36 &    ---    &    ---    &    1.0    \\
AFGL 989		& 29 &    1.7    &    0.8    &    1.6    \\
AFGL 2136		& 77 & {\bf 2.8} & {\bf 1.8} & {\bf 1.7} \\
AFGL 2591		& 50 & {\bf 2.6} &    ---    & {\bf 2.0} \\
AFGL 2789		& 13 &    ---    &    ---    &    2.3    \\
AFGL 2884		& 70 & {\bf 2.6} &    ---    &    0.8    \\
AFGL 4176		& 72 &    ---    &    ---    &    0.8    \\
M8E IR			& 22 &    ---    &    ---    &    0.8    \\
Mon R2 IRS2 	& 70 & {\bf 2.2} & {\bf 2.0} & {\bf 1.9} \\
NGC 7538 IRS1	&117 &    1.1    &    ---    &    1.0    \\
OMC--1 BN		& 58 & {\bf 3.8} & {\bf 3.2} & {\bf 3.9} \\
OMC--1 IRc4		& 65 &    ---    &    ---    &    1.2    \\
Oph 29		    & 30 & {\bf 2.9} &    0.8    &    2.1    \\
SVS 13			& 23 &    ---    & {\bf 3.2} & {\bf 3.8} \\
W3 IRS5			&117 & {\bf 1.4} &    ---    &    1.0    \\
W33A			&120 & {\bf 1.5} &    1.0    & {\bf 1.6} \\
W51 IRS2		& 22 &    ---    &    ---    & {\bf 5.0} \\
\enddata 
\tablenotetext{a}{Values significantly enhanced relative to the trend for Taurus field stars are in bold.}
\tablenotetext{b}{Values of $\ratlam$ for ices refer to data for the $\lambda = 4.67$\mic\ CO feature in W33A, and to the $\lambda = 3.0$\mic\ \water\ feature in all other objects.}
\end{deluxetable}

\begin{deluxetable}{ll}
\tabletypesize{\scriptsize} 
\tablecaption{Key to plotting symbols used in the figures. \label{symbol}} 
\tablewidth{0pt} 
\tablehead{\colhead{Symbol} & \colhead{Meaning}}
\startdata 
Circle				& Photometric colors (Fig.\,1 only) \\
Square  			& Broadband $K$ (2.2\mic)$^a$ \\
Upright triangle	& \water-ice (3.0\mic) \\
Inverted triangle	& CO-ice (4.67\mic) \\
Diamond 			& Silicate (9.7\mic) \\
\\
Filled symbol (any)	& Background field star \\
Open symbol (any)	& Embedded YSO \\
\enddata
\tablenotetext{a}{Also $\lmax$, $\pmax$ in Figs. 9--10.}
\end{deluxetable}

\clearpage

\begin{figure}
    \centering
    \includegraphics[width=12.5cm, angle=0]{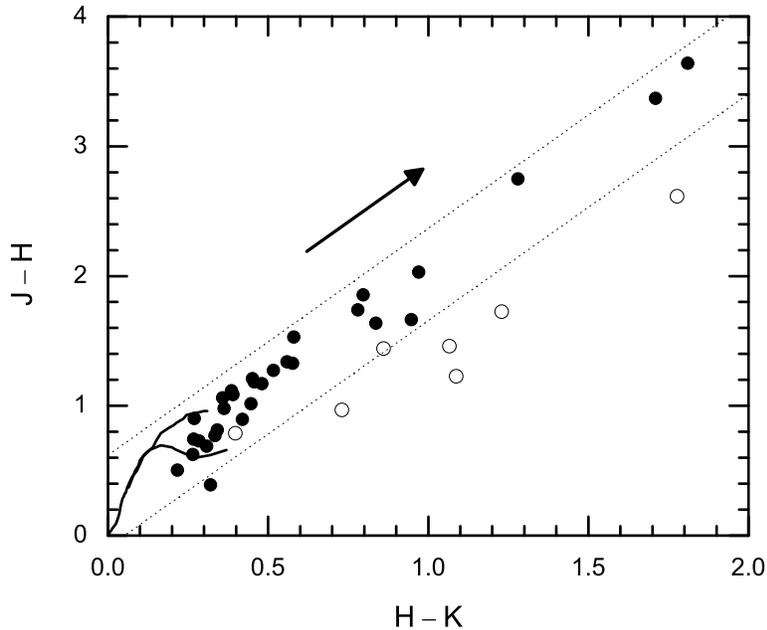}
\caption{{\it JHK\/} color-color diagram for Taurus stars (Table~1) with extinction estimates based on infrared data ($\ejk$ method). Filled circles denote field stars, open circles denote YSOs. Photometry is from the Two-Micron All-Sky Survey (2MASS). Solid curves near the origin are intrinsic colors for normal unreddened stars of luminosity classes III and V (upper and lower branches). The arrow is a sample reddening vector with slope $\ejh/\ehk = 1.75$ (Whittet \etal\ 2007) and amplitude equivalent to $\av\approx 5$~mag. The dotted diagonal lines delimit the expected loci of stars with normal intrinsic colors, subject to reddening. The YSOs are subject to additional displacement resulting from circumstellar infrared emission (e.g.\ Itoh \etal\ 1996), but nevertheless show a general trend roughly parallel to the reddening vector. \label{fig1}}
\end{figure}

\begin{figure}
    \centering
    \includegraphics[width=8.15cm, angle=0]{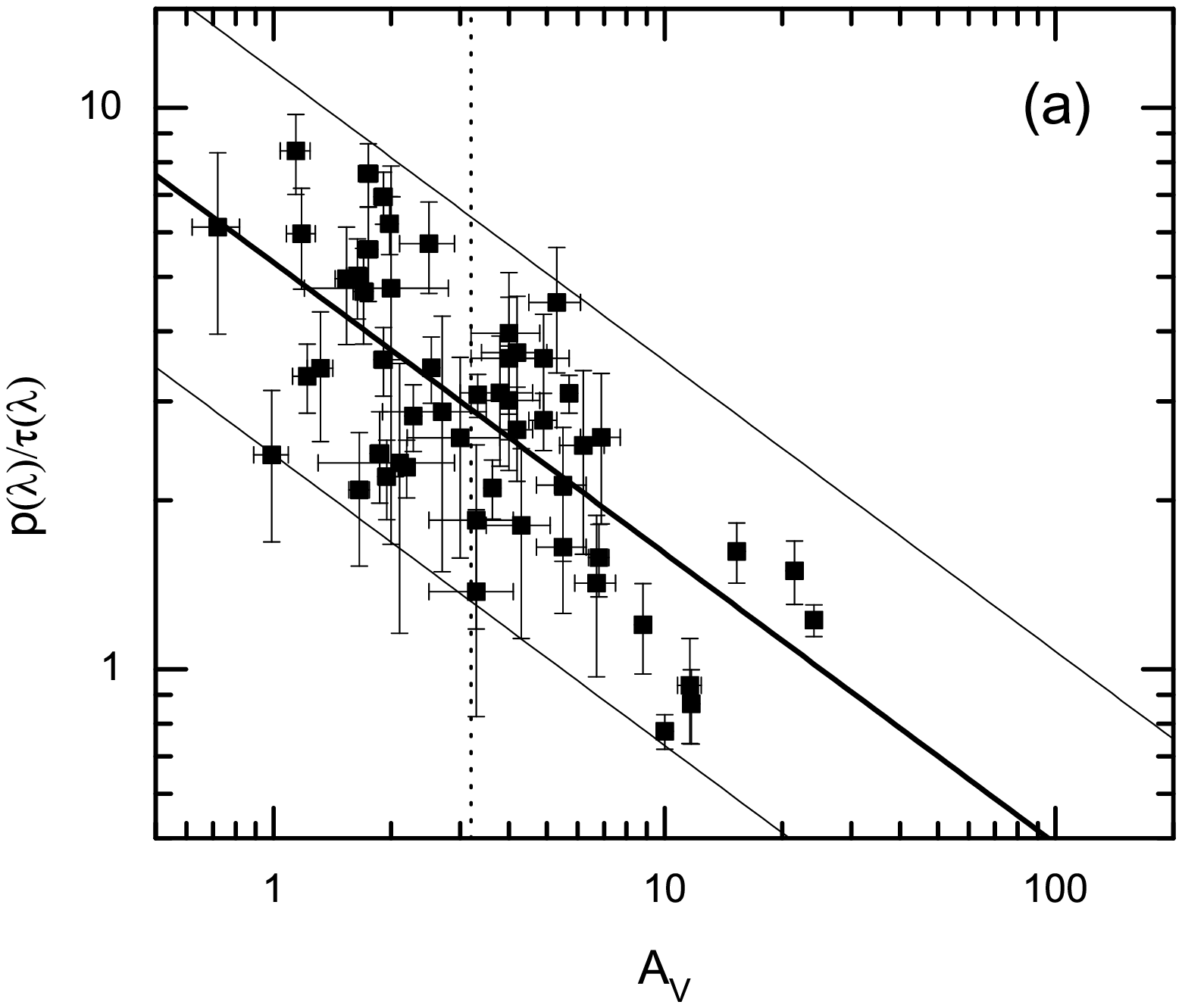}
    \includegraphics[width=8.15cm, angle=0]{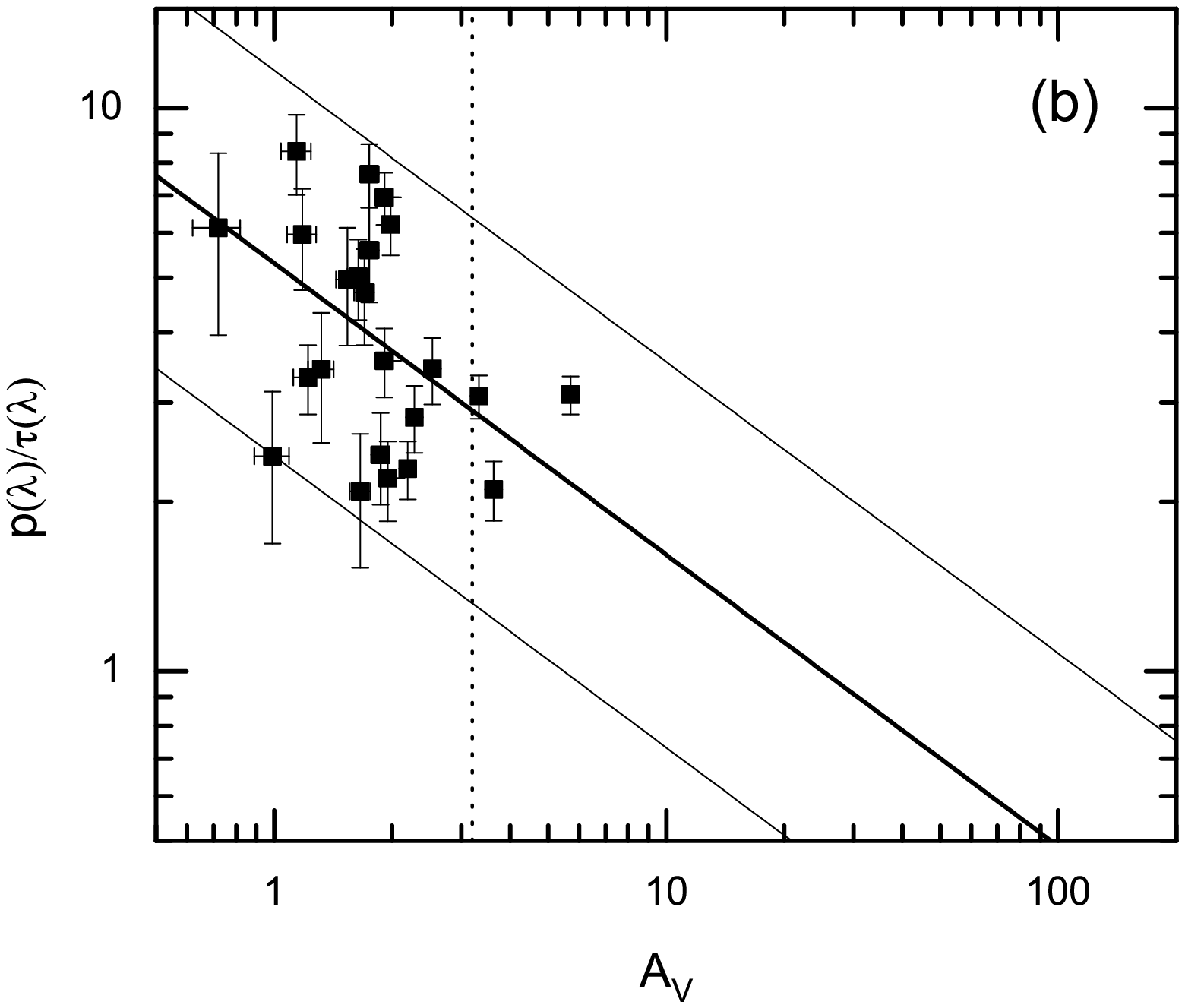}
    \includegraphics[width=8.15cm, angle=0]{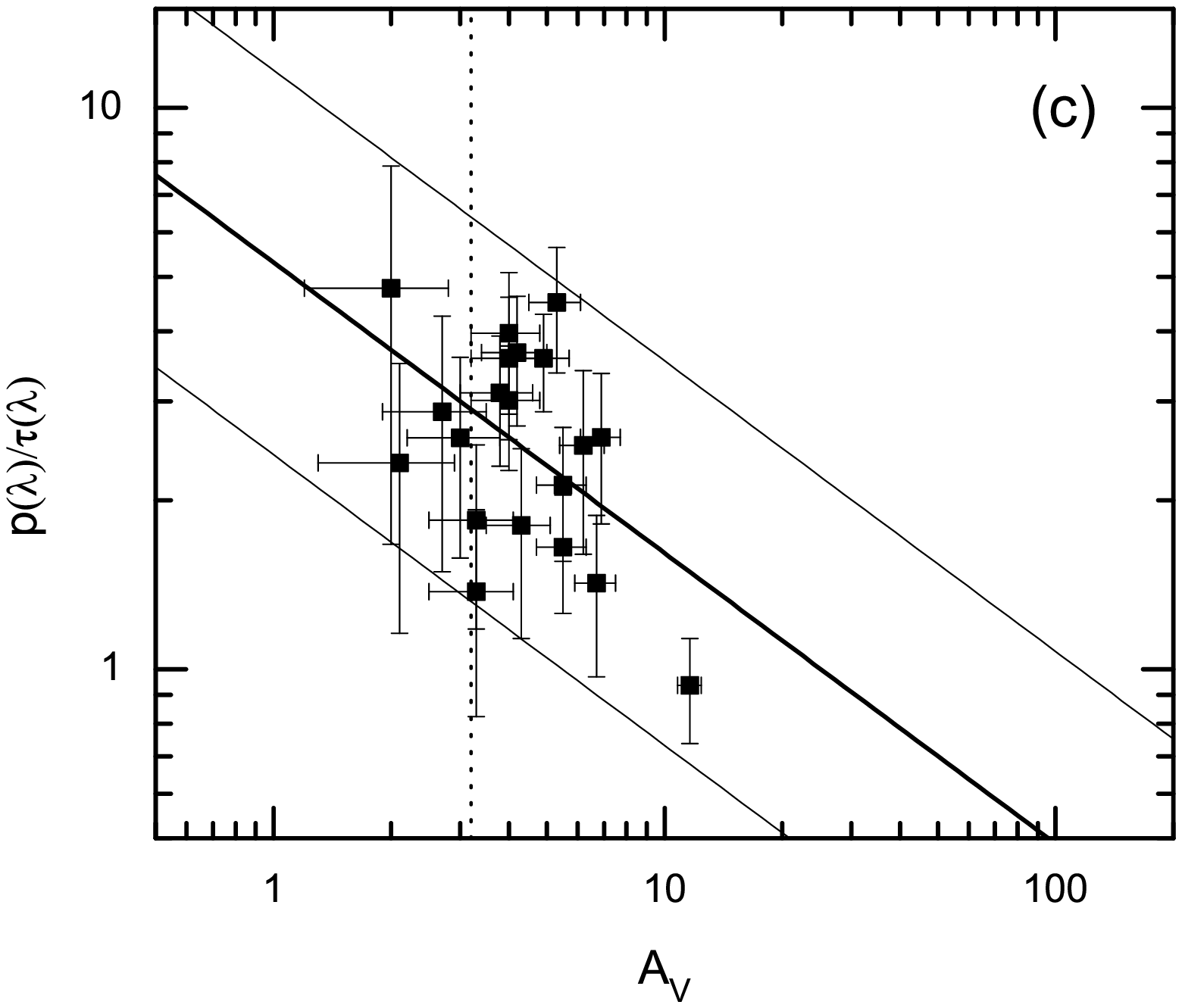}
    \includegraphics[width=8.15cm, angle=0]{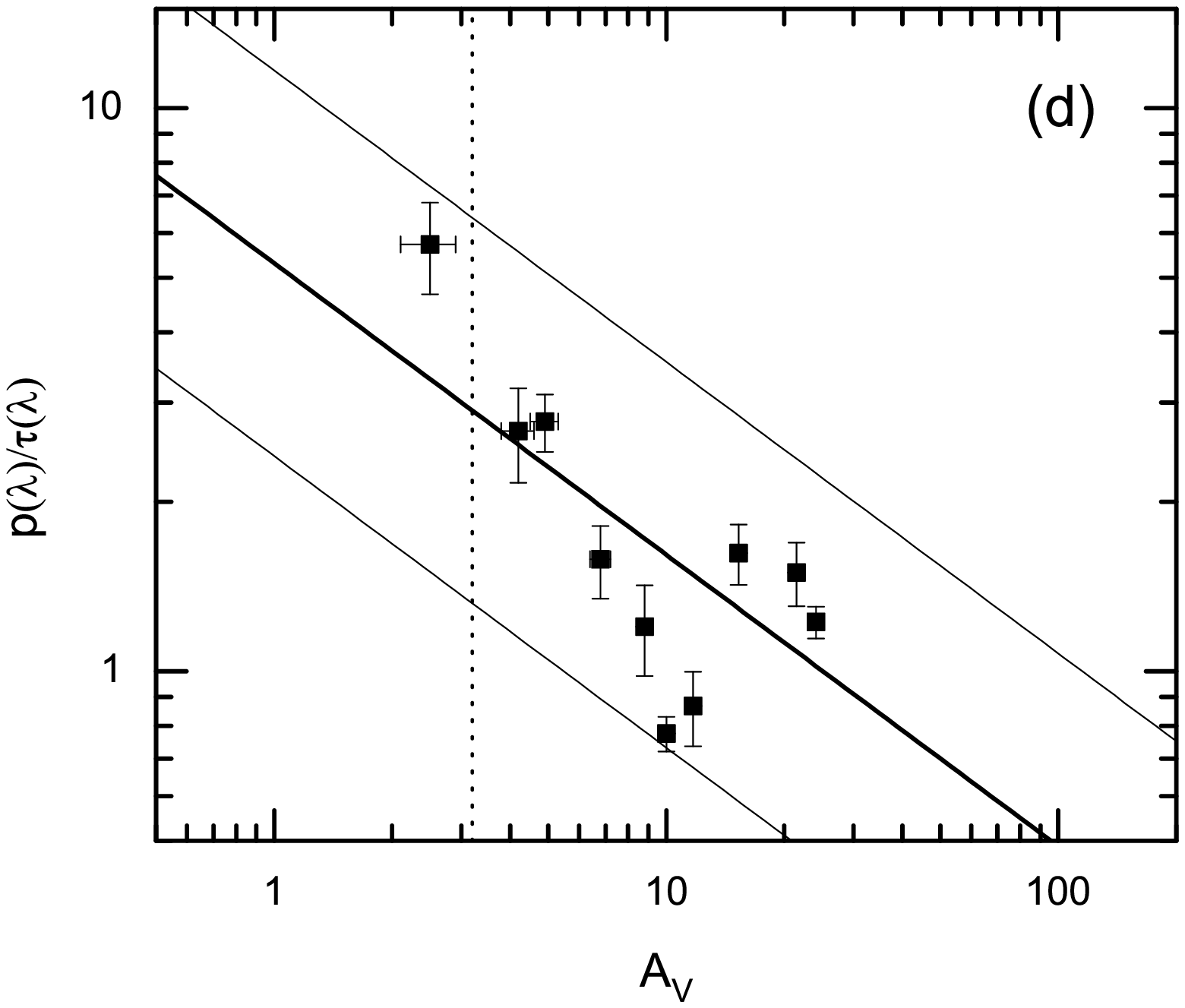}
    \includegraphics[width=8.15cm, angle=0]{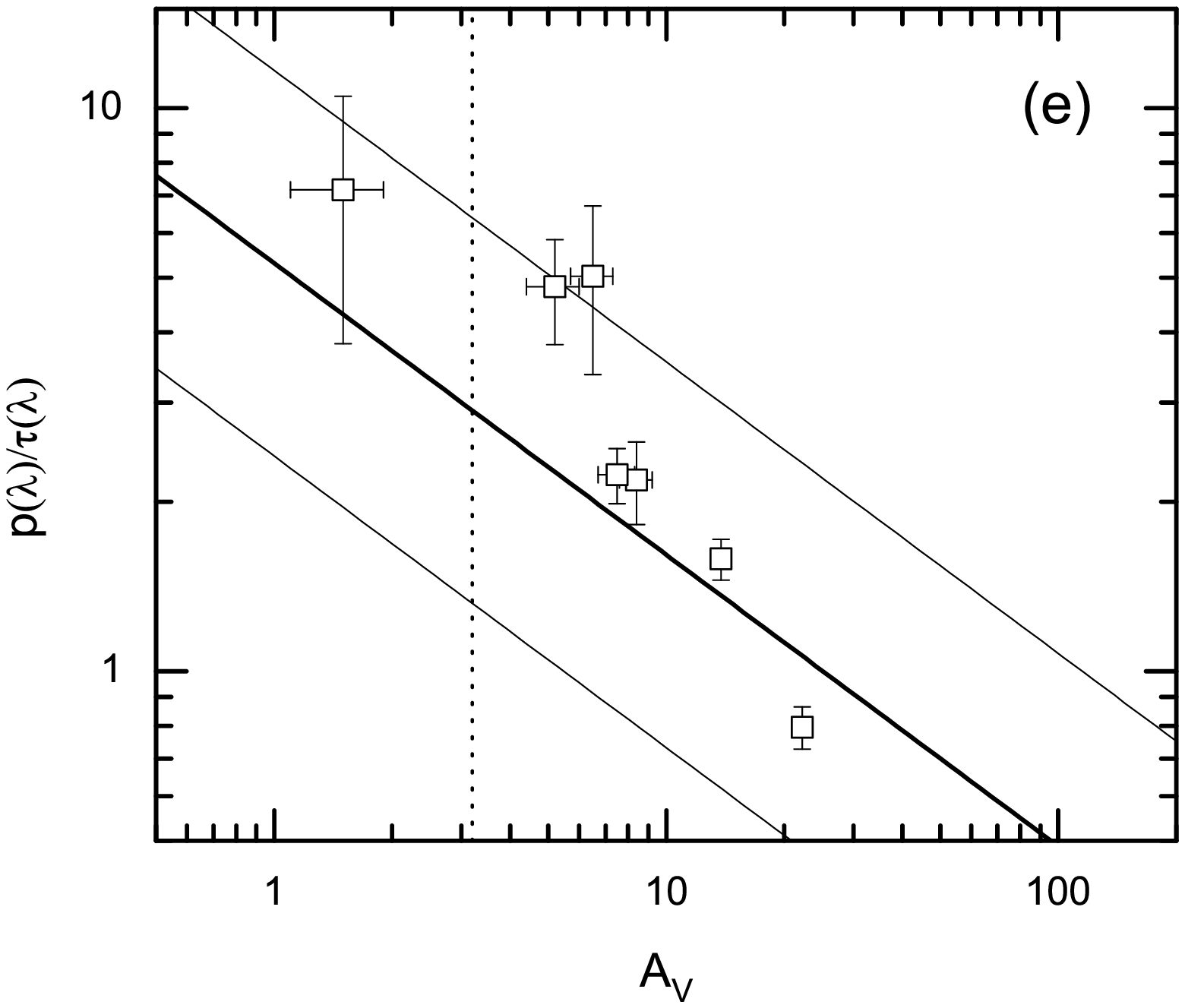}
    \includegraphics[width=8.15cm, angle=0]{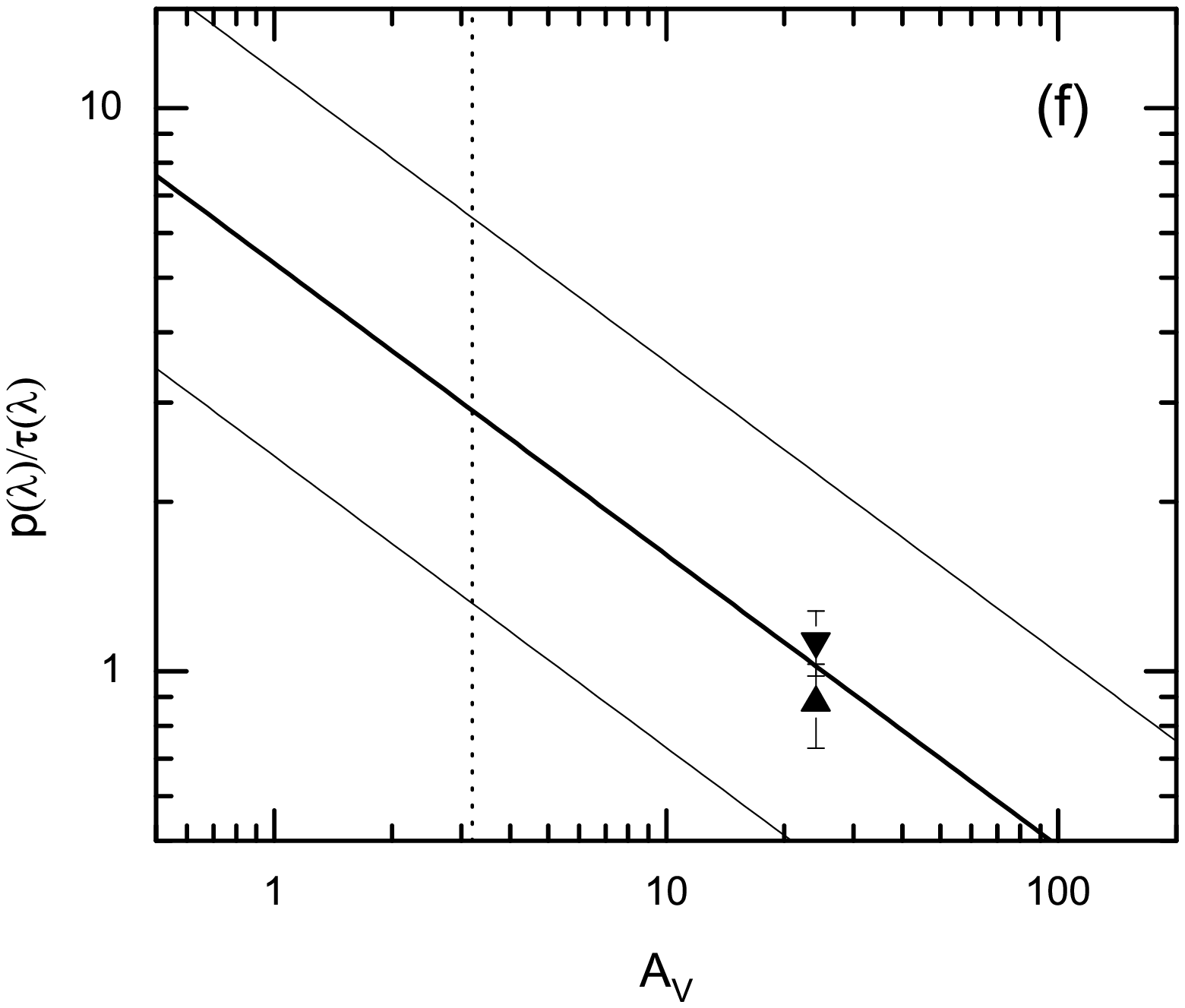}
\end{figure}
\clearpage
\begin{figure}\caption{Plots of polarization efficiency ($\ratlam$) vs.\ visual extinction ($\av$) for Taurus stars. Frames (a) through (d) show 2.2\mic\ broadband data for field stars (Table~1): (a)~the entire sample; (b)~subset with $\av$ from the $\evk$ method; (c)~subset with $\av$ from the $\ejk$ method (no spectral types); (d)~subset with $\av$ from the $\ejk$ method (with spectral types). Frame~(e) plots 2.2\mic\ broadband data for Taurus YSOs (Table~1). Frame~(f) plots values based on 3.0\mic\ (\water) and 4.67\mic\ (CO) ice-feature spectropolarimetry (Table~2) for the Taurus field star Elias~16. The thick diagonal line in each frame is the least-squares power-law fit to the 2.2\mic\ data for all field stars (as in frame~a), and the thinner diagonal lines represent the approximate upper and lower bounds of the distribution ($\pm 0.34$~dex relative to the power-law fit). The vertical dotted line in each frame denotes the ``threshold extinction" for \water-ice in the cloud ($\av=3.3$~mag; Whittet \etal\ 2001), above which the grains typically acquire mantles. \label{fig2}}
\end{figure}
\clearpage

\begin{figure}
\centering
\includegraphics[width=12.5cm, angle=0]{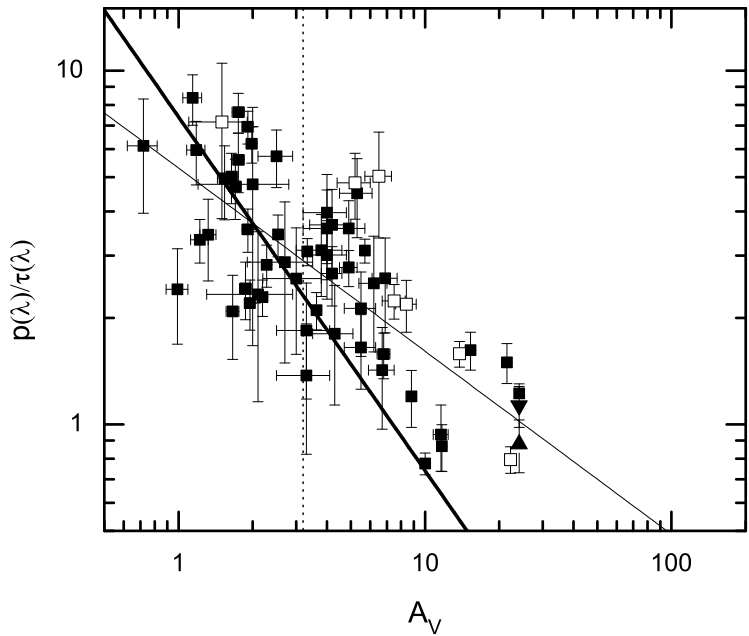}
\caption{Taurus data compared with a simple ``skin-depth" model for the variation of polarization efficiency with extinction. All data plotted in \fig 2 are included here in a single frame; the plotting symbols are explained in Table~\ref{symbol}. The thin diagonal line is the fit to 2.2\mic\ field-star data, as in \fig 2, and the vertical dotted line denotes the ice threshold extinction. The thick diagonal line assumes that $p(\lambda)/\tau(\lambda)\propto[\av]^{-1}$, as expected if polarization is effectively constant, variations in $p(\lambda)/\tau(\lambda)$ with $\av$ resulting merely from the increase in optical depth. \label{fig3}}
\end{figure}
\clearpage

\begin{figure}
    \centering
    \includegraphics[width=12.5cm, angle=0]{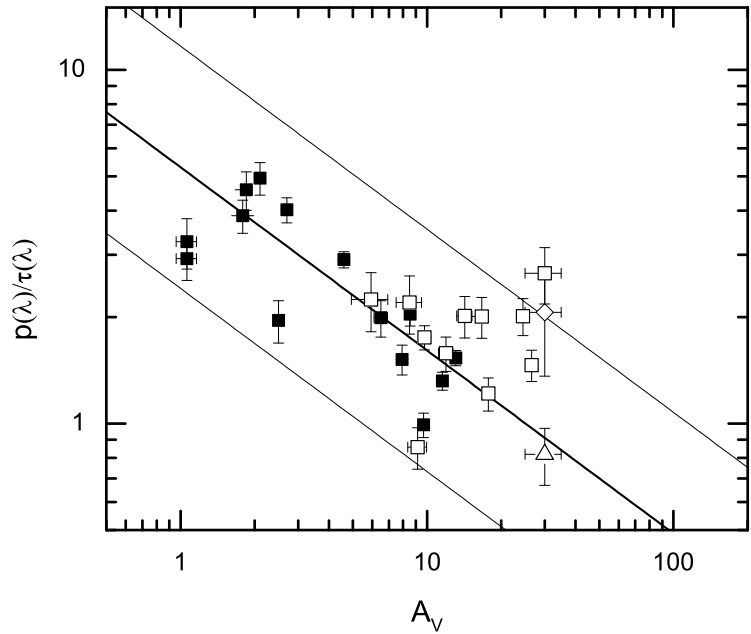}
\caption{Polarization efficiency vs.\ visual extinction for the \oph\ dark cloud. The plotting symbols are explained in Table~\ref{symbol}. $K$-band data are from Table~\ref{oph}; \water-ice and silicate data for YSO Oph~29 ($\av\approx 30$~mag) are from sources listed in Table~\ref{other}. The diagonal lines are identical to those in \fig 2, i.e., the power-law fit and upper/lower bounds to the distribution of field-star data for the Taurus cloud. \label{fig4}}
\end{figure}

\begin{figure}
    \centering
    \includegraphics[width=12.5cm, angle=0]{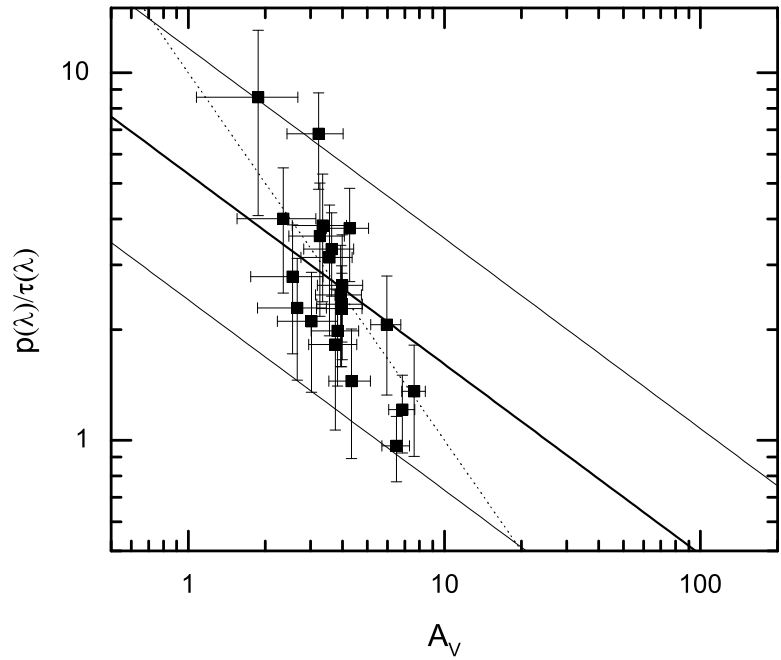}
\caption{Polarization efficiency vs.\ visual extinction for field stars observed through the filamentary dark cloud L1755 (see Table~\ref{other}). The solid diagonal lines are identical to those in \fig 2, i.e., the power-law fit and upper/lower bounds to the distribution of field-star data for the Taurus cloud. The dotted line is a $p(\lambda)/\tau(\lambda)\propto[\av]^{-1}$ model, as in \fig 3. \label{fig5}}
\end{figure}

\begin{figure}
    \centering
    \includegraphics[width=12.5cm, angle=0]{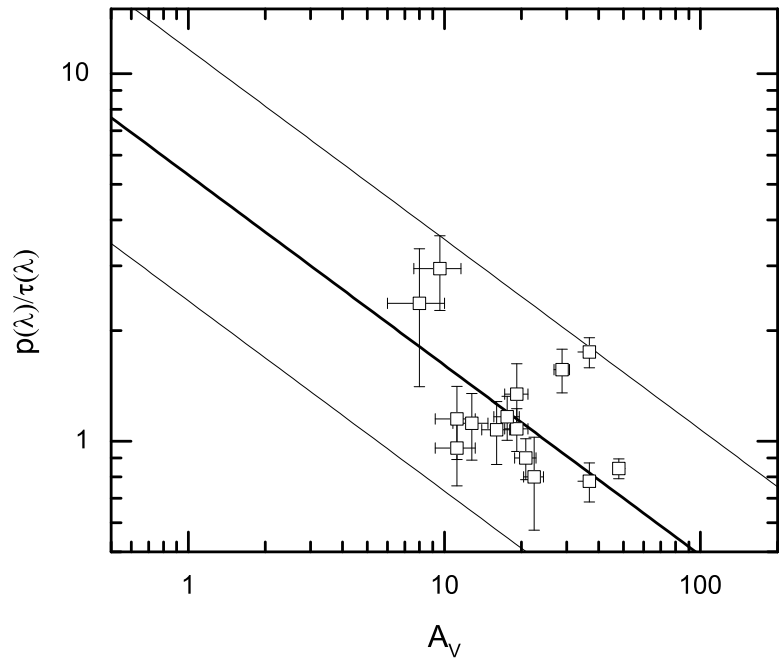}
\caption{Polarization efficiency vs.\ visual extinction for YSOs in the L1641 dark cloud (see Table~\ref{other}). Diagonal lines represent the power-law fit and upper/lower bounds to the distribution of data for Taurus field stars, as in \fig 2.  \label{fig6}}
\end{figure}

\begin{figure}
    \centering
    \includegraphics[width=9.5cm, angle=0]{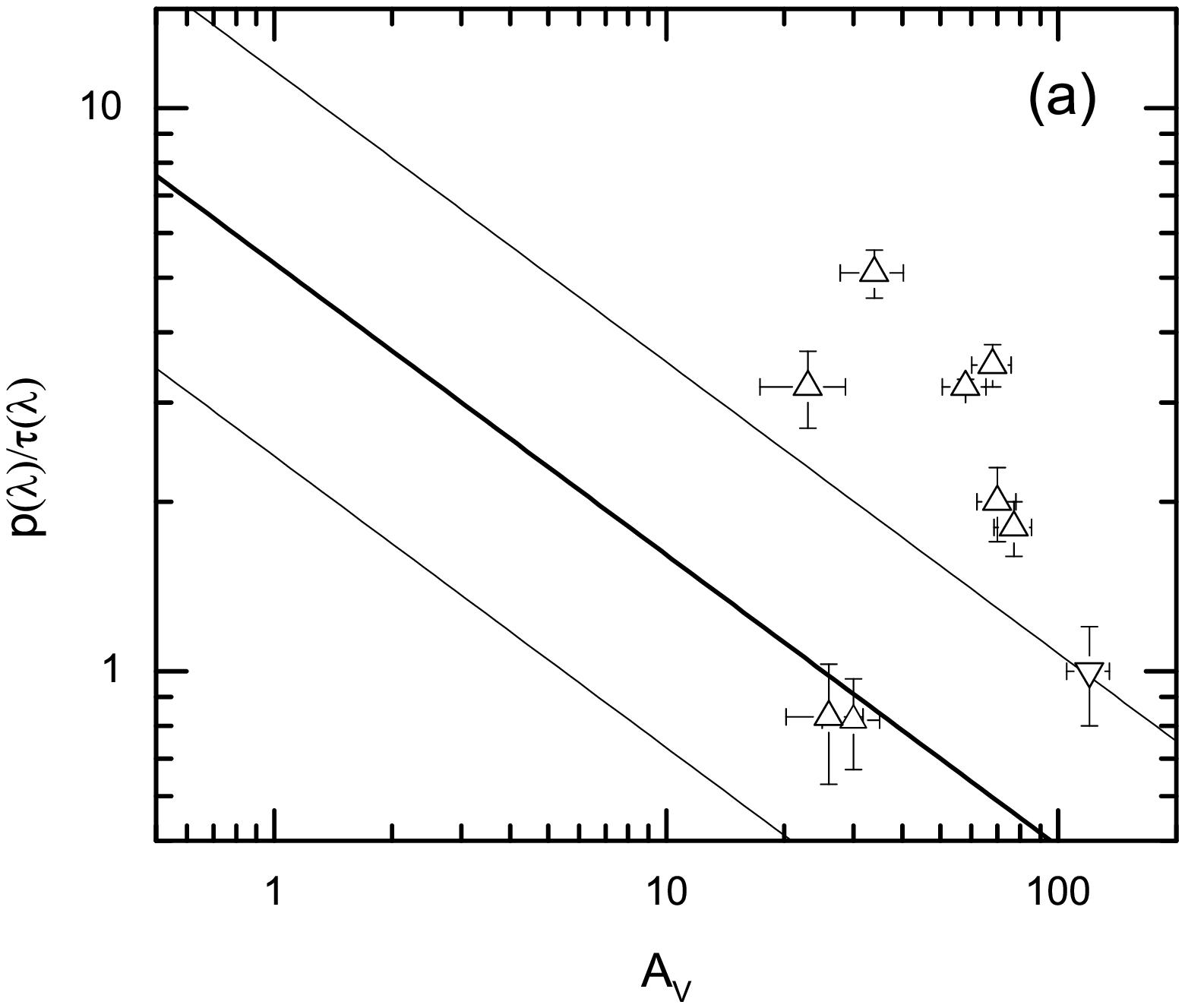}
    \includegraphics[width=9.5cm, angle=0]{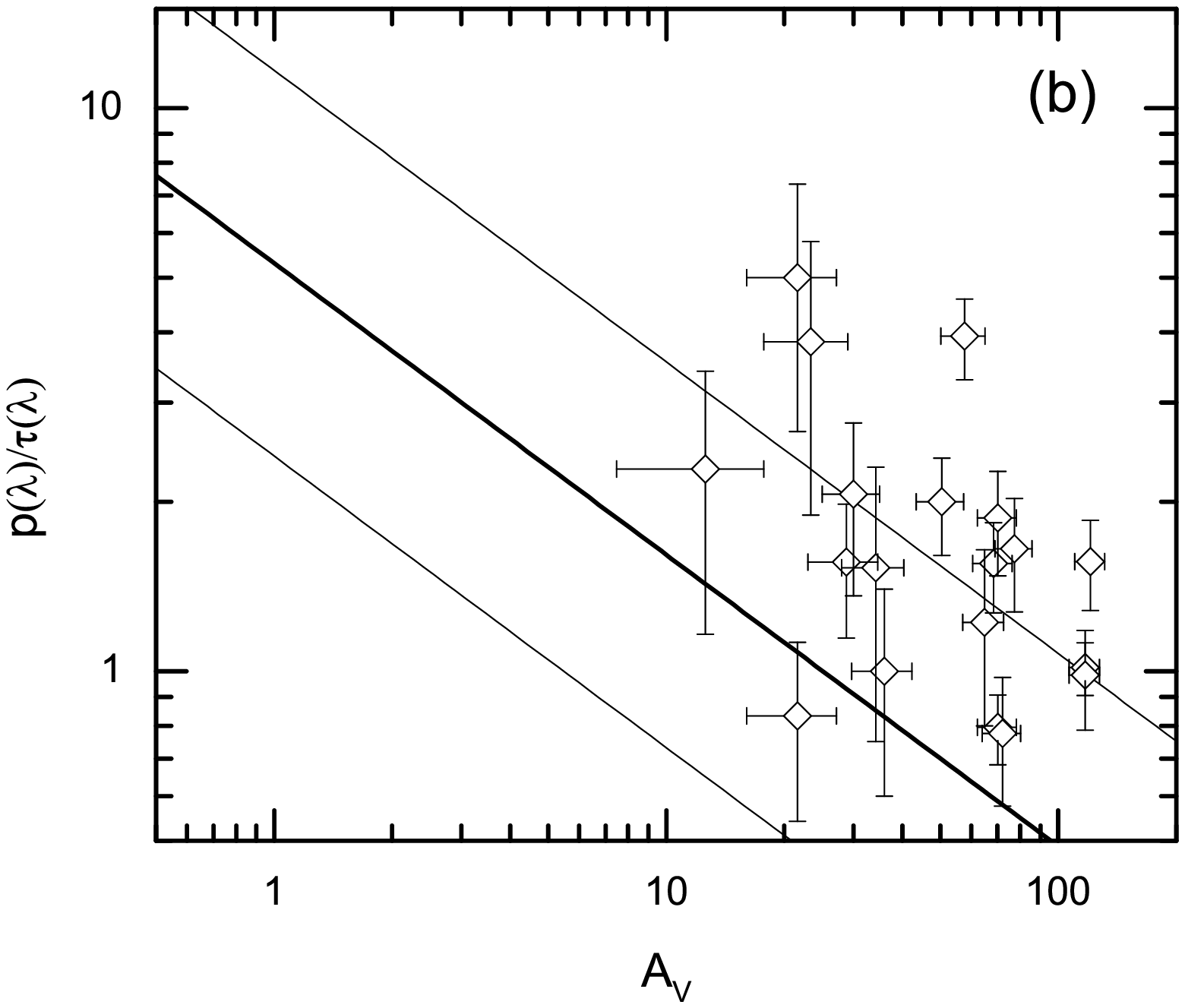}
    \includegraphics[width=9.5cm, angle=0]{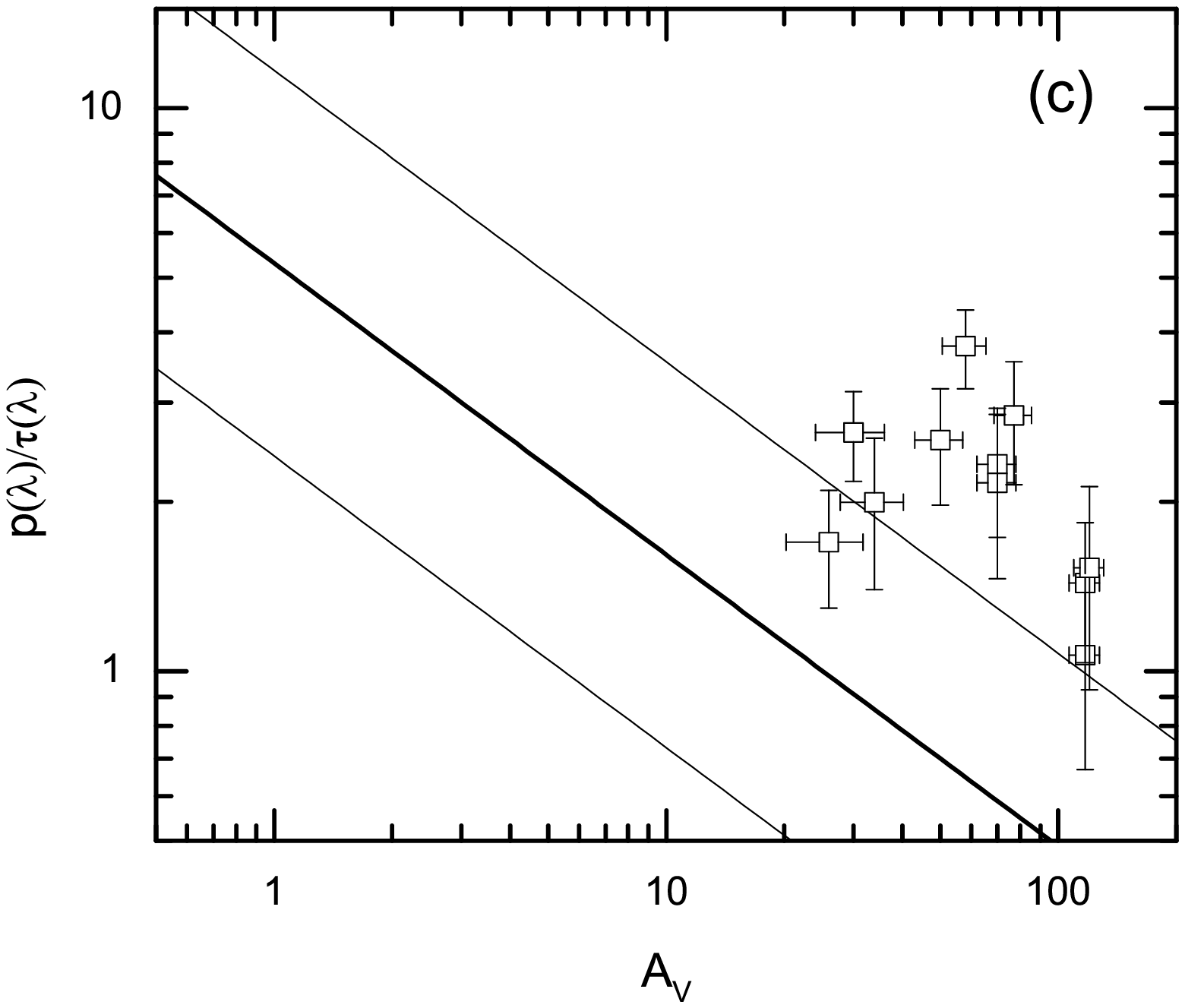}
\end{figure}
\clearpage
\begin{figure}\caption{Polarization efficiency vs.\ visual extinction for solid-state features observed in YSOs (see Tables~\ref{other} and \ref{ysos}): \water\ and CO ice data are plotted in frame~(a), silicate data in frame~(b); $K$-band continuum data as available for the same stars are plotted in frame~(c). Diagonal lines represent the power-law fit and upper/lower bounds to the distribution of data for Taurus field stars, as in \fig 2. \label{fig7}}
\end{figure}

\begin{figure}
\centering
\includegraphics[width=11cm, angle=0]{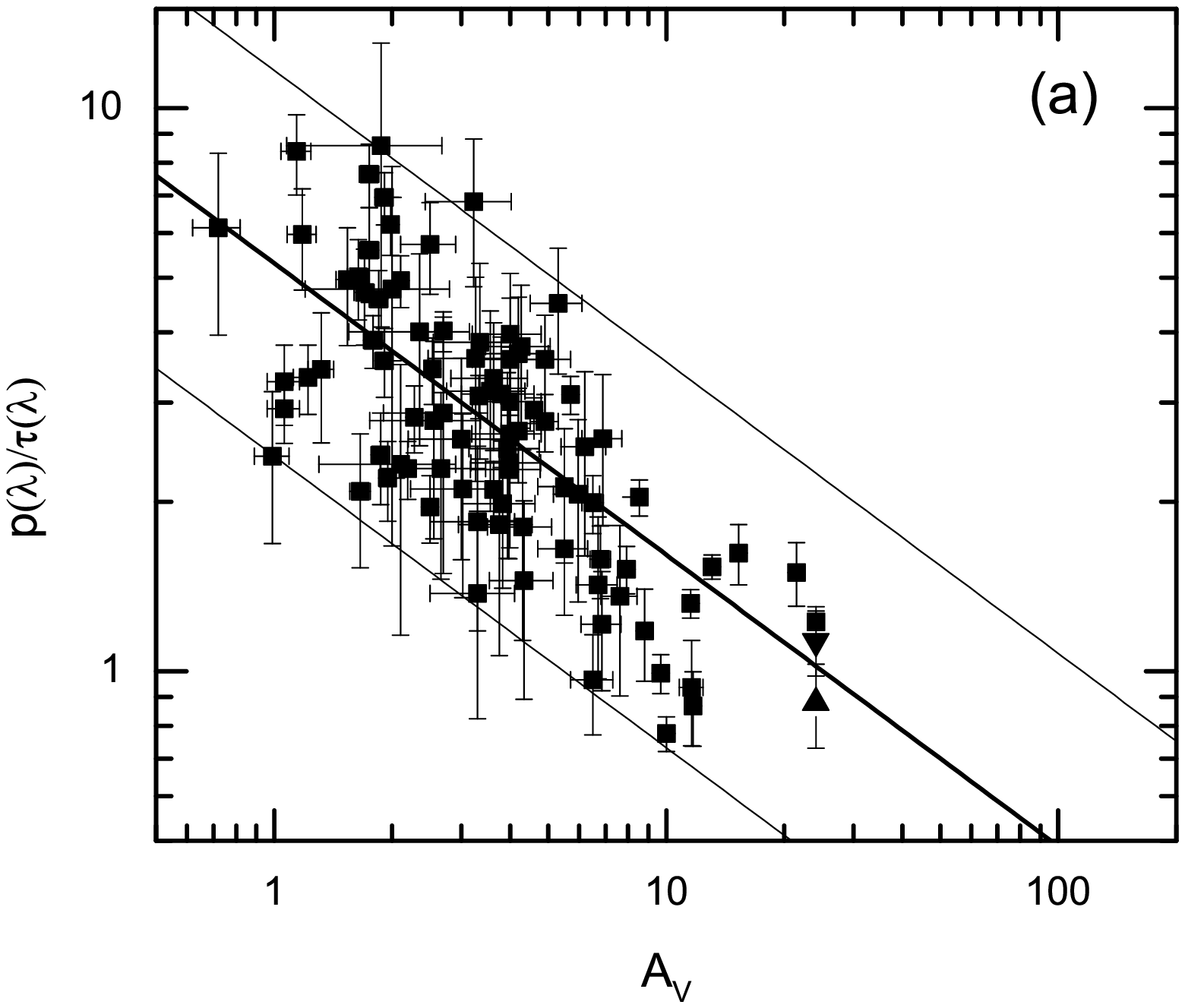}
\includegraphics[width=11cm, angle=0]{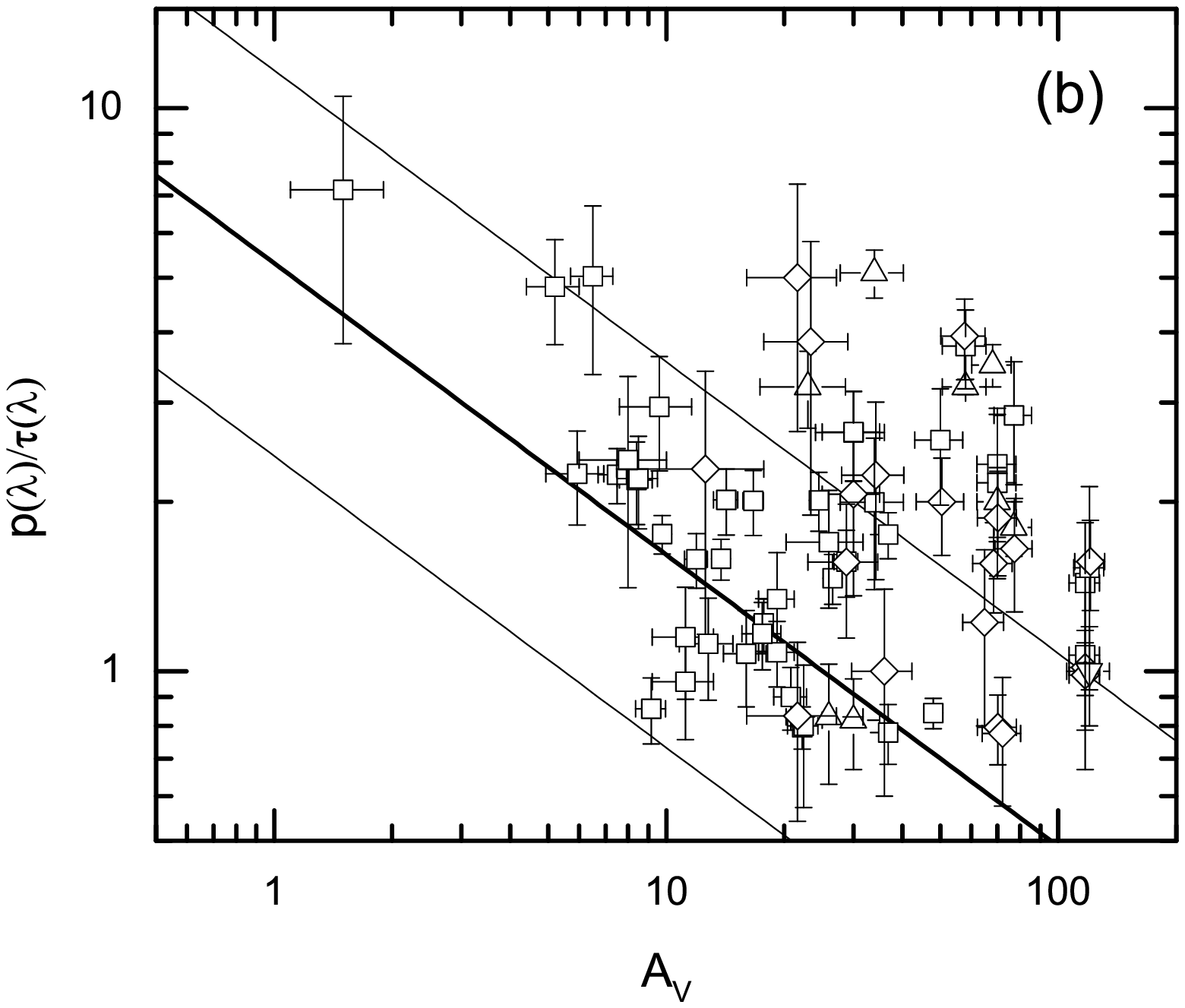}
\caption{Summary plots of all data in Figs.~2--7 (field stars: frame~a; YSOs: frame~b). The plotting symbols are explained in Table~\ref{symbol}, and the diagonal lines are again identical to those in \fig 2. \label{fig8}}
\end{figure}
\clearpage

\begin{figure}
\centering
\includegraphics[width=12.4cm, angle=0]{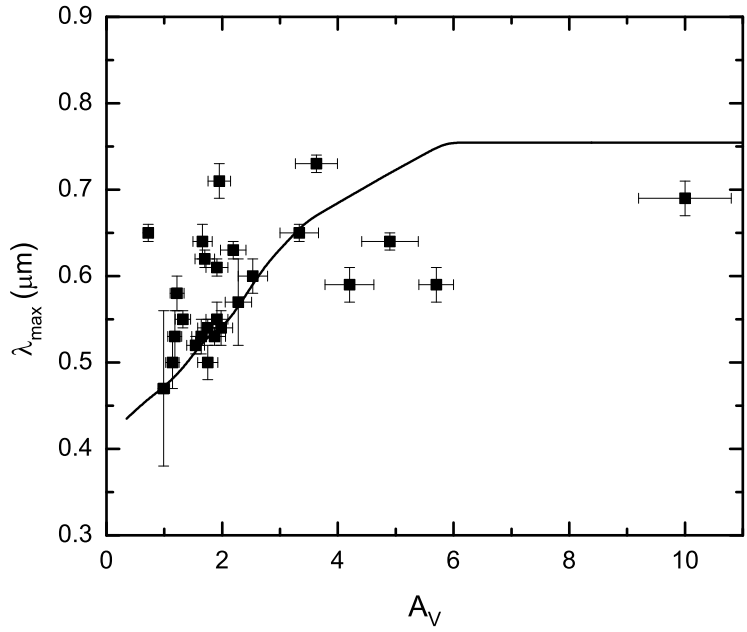}
\caption{Plot of $\lambda_{\rm max}$ vs.\ $\av$, comparing observational data for Taurus field stars (Whittet \etal\ 2001) with a prediction for the radiative torques model described in \S4 (curve). The $\lambda_{\rm max}$ value is effectively a measure of the mean size of the aligned grains. \label{fig9}}
\end{figure}
\clearpage

\begin{figure}
\centering
\includegraphics[width=11cm, angle=0]{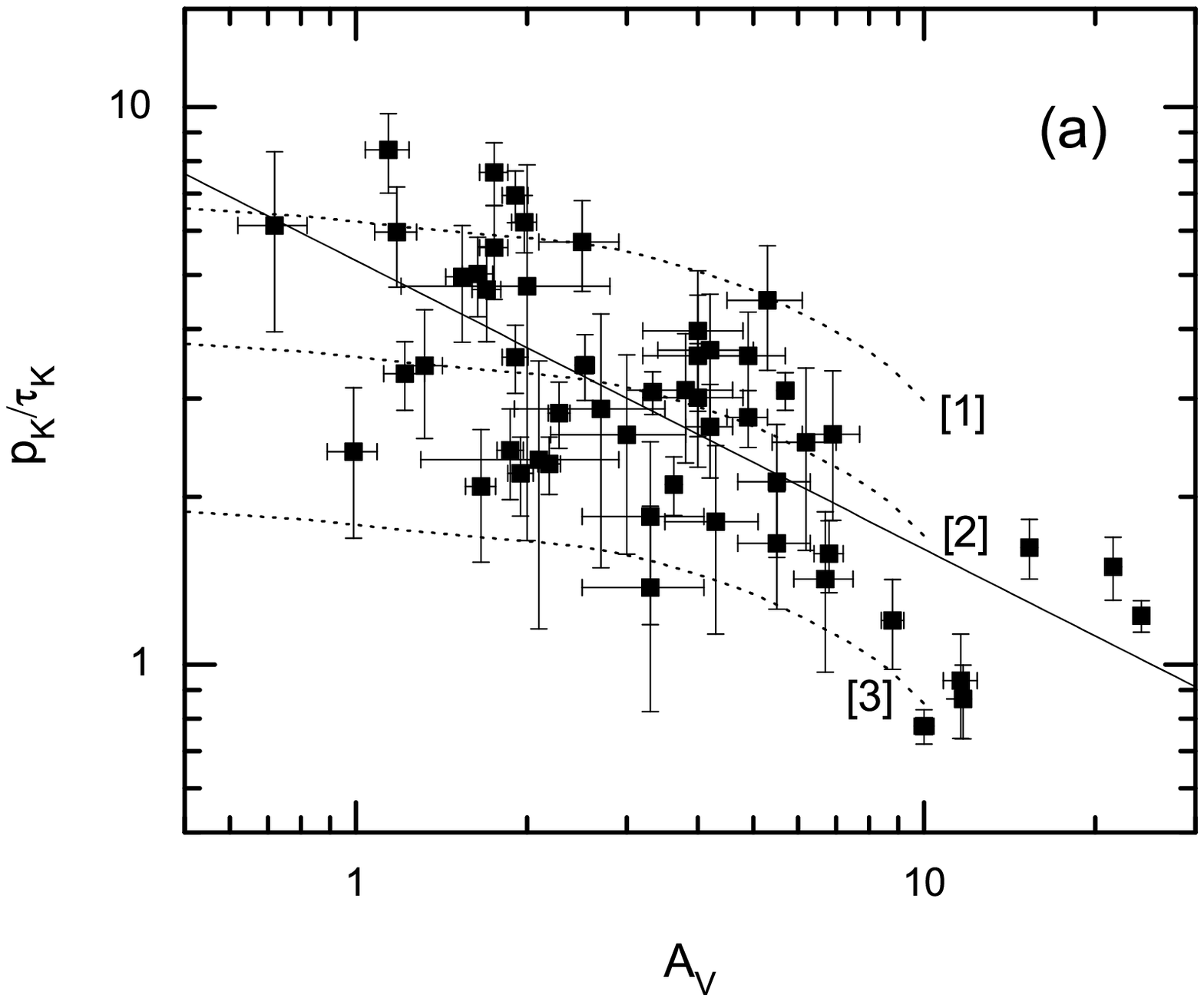}
\includegraphics[width=11cm, angle=0]{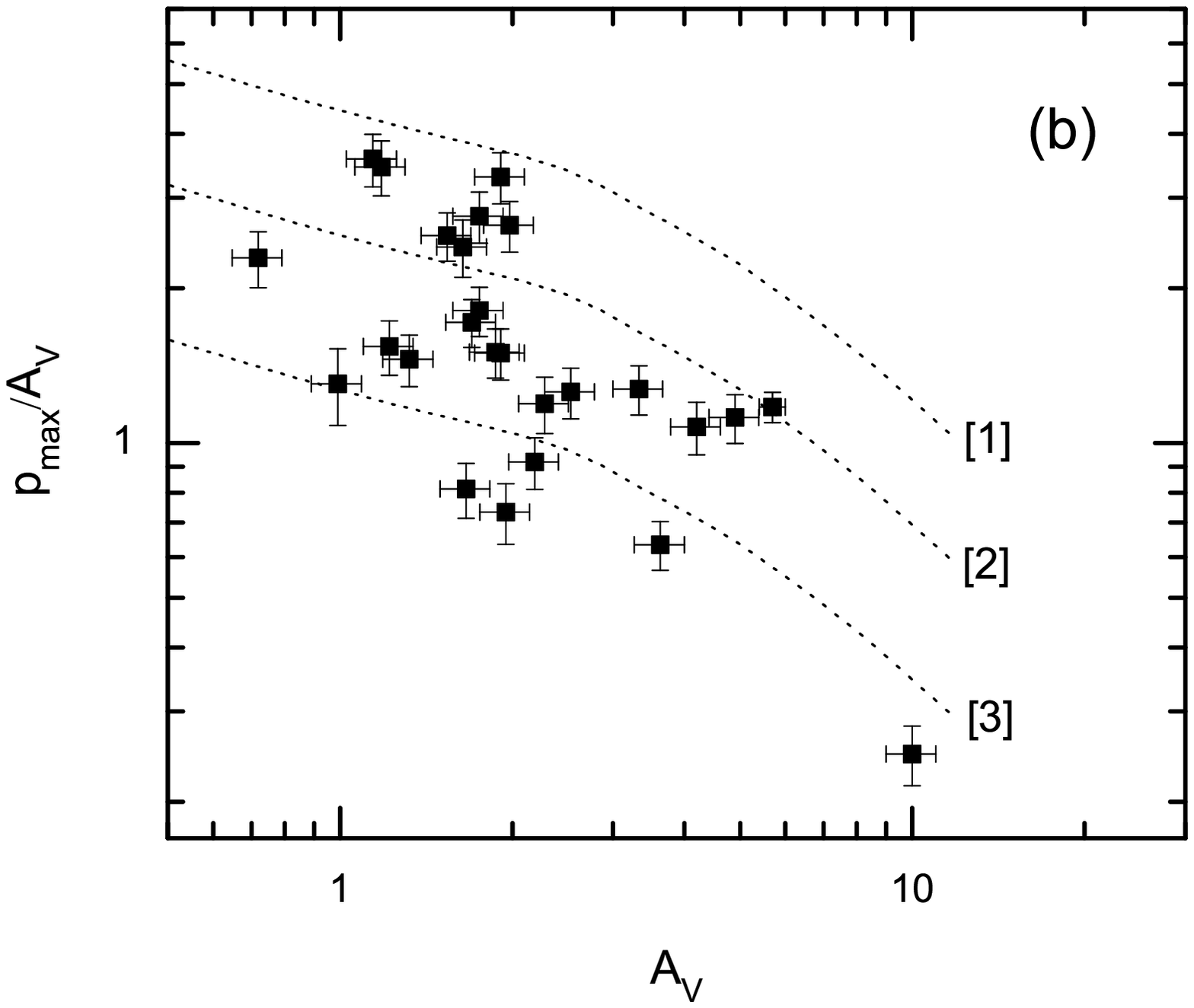}
\caption{Polarization efficiency vs.\ visual extinction, comparing data for Taurus field stars with predictions of the radiative torques model (dotted curves, \S4). Frame~(a): $\ratk$ vs.\ $\av$ (data from Table~1); the diagonal line is the power-law fit to the data, as in \fig 2. Frame~(b): $p_{\rm max}/\av$ vs.\ $\av$ (data from Whittet \etal\ 2001). The curves labeled $[1]$, $[2]$ and $[3]$ in each frame are for $R^{*}$ values (\S4) of 0.7, 0.4 and 0.2, respectively. \label{fig10}}
\end{figure}
\clearpage

\end{document}